\documentclass[aps,prd,groupedaddress,showpacs]{revtex4}

\usepackage{graphicx}
\usepackage{dcolumn}
\usepackage{bm}
\usepackage{amsmath}
\usepackage{epstopdf}
\usepackage{amsfonts}
\usepackage{amssymb}
\usepackage{hyperref}
\usepackage{xcolor}
\usepackage{float}%Ubicacion_figuras

\usepackage{natbib}

\providecommand{\U}[1]{\protect\rule{.1in}{.1in}}

\newcommand{\baa}{\begin{align}}
\newcommand{\eaa}{\end{align}}

\newcommand{\be}{\begin{equation}}
\newcommand{\ee}{\end{equation}}
\newcommand{\bea}{\begin{eqnarray}}
\newcommand{\eea}{\end{eqnarray}}

\begin{document}

\title{Dynamical systems methods and statender diagnostic of 
\\
interacting vacuum energy models}

\author{Grigoris Panotopoulos}
\affiliation{Centro de Astrof{\'i}sica e Gravita{\c c}{\~a}o, Instituto Superior T{\'e}cnico-IST,
Universidade de Lisboa-UL, Av. Rovisco Pais, 1049-001 Lisboa, Portugal}
\email{grigorios.panotopoulos@tecnico.ulisboa.pt}

\author{\'Angel Rinc\'on}
\affiliation{Instituto de F\'isica, Pontificia Universidad Cat\'olica de Valpara\'iso, Avenida Brasil 2950, Casilla 4059, Valpara\'iso, Chile.}
\email{angel.rincon@pucv.cl}

\author{Giovanni Otalora}
\affiliation{Instituto de F\'isica, Pontificia Universidad Cat\'olica de Valpara\'iso, Avenida Brasil 2950, Casilla 4059, Valpara\'iso, Chile.}
\email{giovanni.otalora@pucv.cl}

\author{Nelson Videla}
\affiliation{Instituto de F\'isica, Pontificia Universidad Cat\'olica de Valpara\'iso, Avenida Brasil 2950, Casilla 4059, Valpara\'iso, Chile.}
\email{nelson.videla@pucv.cl}

\date{\today}

\begin{abstract}
We study three interacting dark energy models within the framework of four-dimensional General Relativity and a spatially flat Universe. In particular, we first consider two vacuum models where dark energy interacts with dark matter, while relativistic matter as well as baryons are treated as non-interacting fluid components. Secondly, we investigate a third model where the gravitational coupling is assumed to be a slowly-varying function of the Hubble rate and dark energy and dark matter interact as well. We compute the statefinders parameters versus red-shift as well as the critical points and their nature applying dynamical systems methods. In the case of only an interaction term, our main findings indicate that i) significant differences between the models are observed as we increase the strength of the interaction term, and ii) all the models present an unique attractor corresponding to acceleration. On the other hand, when we allow for a variable gravitational coupling, we find that i) the deviation from the concordance model depends of both the strength of gravitational coupling parameter and the interaction term, and ii) there is an unique attractor corresponding to acceleration. 
\end{abstract}

\pacs{}

\maketitle

%%%%%%%%%%%%%%%%%%%%%%%%%
\section{Introduction}
%%%%%%%%%%%%%%%%%%%%%%%%%

The origin and nature of dark energy (DE), the fluid component that currently accelerates the Universe \cite{SN1,SN2,turner}, is one of the biggest mysteries and challenges in modern theoretical Cosmology. Clearly, Einstein's General Relativity \cite{GR} with radiation and matter only cannot lead to accelerating solutions. A positive cosmological constant \cite{einstein} is the simplest, most economical model in a very good agreement with a great deal of current observational data. Since, however, it suffers from the cosmological constant (CC) problem \cite{weinberg}, other possibilities have been considered in the literature over the years. The CC problem, introduced by Zeldovich for the first time more than fifty years ago \cite{zeldovich}, may be summarized in a few words as follows: It is an impressive mismatch-by many orders of magnitude-between the observational value of vacuum energy, and the expected value from particle physics due to vacuum fluctuations of massive fields. Although some progress has been made up to now, see e.g. \cite{Garriga:2000cv,Padmanabhan:2013hqa,Mikovic:2014opa,Canales:2018tbn}, the origin of the CC problem still remains a mystery.

Regarding the CC problem and possible alternatives to the $\Lambda$CDM model, either a modified theory of gravity is assumed, providing correction terms to GR at cosmological scales, or a new dynamical degree of freedom with an equation-of-state (EOS) parameter $w < -1/3$ must be introduced. In the first class of models (geometrical DE) one finds for instance $f(R)$ theories of gravity \cite{mod1,mod2,HS,starobinsky}, brane-world models \cite{langlois,maartens,dgp} and Scalar-Tensor theories of gravity \cite{BD1,BD2,leandros,PR}, while in the second class (dynamical DE) one finds models such as quintessence \cite{DE1}, phantom \cite{DE2}, quintom \cite{DE3}, tachyonic \cite{DE4} or k-essence \cite{DE5}. For an excellent review on the dynamics of dark energy see e.g. \cite{copeland}.

Furthermore, regarding the value of the Hubble constant $H_0$, there is nowadays a tension between high red-shift CMB data and low red-shift data, see e.g. \cite{tension,tension1,tension2,tension3}. The value of the Hubble constant extracted by the PLANCK Collaboration \cite{planck1,planck2}, $H_0 = (67-68)~\text{km/(Mpc  sec)}$, is found to be lower than the value obtained by local measurements, $H_0 = (73-74)~\text{km/(Mpc sec)}$ \cite{hubble,recent}. This tension might call for new physics \cite{newphysics}.
What is more, regarding large scale structure formation data, the growth rate from red-shift space distortion measurements has been found to be lower than expected from PLANCK \cite{eriksen,basilakos}. 

Both tensions may be alleviated within the framework of running vacuum dynamics  \cite{sola1,sola2,sola3,sola4,sola5,sola6,sola7}. In this class of models, contrary to a rigid cosmological constant $\Lambda = const.$, vacuum energy density can be expressed as a function
of the Hubble rate, i.e. $\rho_{\Lambda}=\rho_{\Lambda}(H)$, being of dynamical nature and at the same time it may interacts with dark matter, and also it accounts for a running of the gravitational coupling $G$ \cite{Fritzsch:2016ewd}. We remark in passing that other alternatives approaches to running vacuum dynamics  do exist, and one may mention for instance the scale--dependent (SD) scenario \cite{Reuter:2003ca,Koch:2016uso,Hernandez-Arboleda:2018qdo}, in which it is assumed that the couplings of the original classical action acquire a scale--dependence. The SD scenario is one of the approaches to quantum gravity, inspired by the well--known Brans--Dicke theory \cite{BD1,BD2}, where Newton's constant is replaced by an dynamical scalar field following the identification $\phi \rightarrow G^{-1}$. In SD cosmological models the cosmological constant becomes time dependent similarly to the running vacuum dynamics, although in the SD scenario Newton's constant, too, acquires a time dependence.

Remarkably, measurements of the expansion rate based on Hubble-diagram of high-redshift objects \cite{Riess:2019cxk,Risaliti:2018reu} suggest that  a  rigid $\Lambda$ term is  ruled  out  by a statistical significance of  $\sim 4\sigma$, accounting for deviations from $\Lambda$CDM model.
The aforementioned deviations allow the possibility of both dynamical and interacting DE, which is realizable within the framework of running vacuum scenario \cite{sola7}. More generically, interacting DE models are interesting for several reasons. First of all, it is a possibility that should not be ignored, under the assumption
that DE and DM do not evolve separately but interact with each other non-gravitationally. 
Secondly, and perhaps the main motivation for
an interaction in the dark
sector, is currently motivated that this scenario can solve the current
cosmological tensions in some data, see e.g. \cite{Kumar:2016zpg,Kumar:2017dnp,Yang:2018euj}. Additionally, other recent relevant results regarding the interaction between DE and DM were found in \cite{Kumar:2017bpv,Yang:2019vni}. For an extensive review on DE and DM interactions, see \cite{Wang:2016lxa} an references therein. In addition, the "why now problem" may be addressed if our current Universe sits at a stable fixed point (attractor) of the corresponding dynamical system, and this attractor corresponds to acceleration and to $0 < \Omega_{m,0} < 1$, with $\Omega_{m,0}$ being today's normalized density of matter. Thus, the system will always reach its attractor at late times irrespectively of the initial conditions. It can be easily shown that this scenario cannot be realized if there is no interaction between DE and matter \cite{KPT}.

As several DE models predict very similar expansion histories, all of them are still in agreement with the available observational data. It thus becomes clear that it is advantageous to introduce and study new appropriate quantities capable of discriminating between different dark energy cosmological models at least at background level. 
Hence, in order to compare different dark energy models we can introduce parameters in which derivatives of the scale factor beyond the second-order appear.
To this end, one option would be to study the so-called statefinder parameters, $r,s$, defined as follows \cite{Sahni:2002fz,Alam:2003sc}
\begin{eqnarray}
r & \equiv & \frac{\dddot{a}}{a H^3}, \\
s & \equiv & \frac{r-1}{3 (q-\frac{1}{2})} \label{rsz},
\end{eqnarray}
where the dot denotes differentiation with respect to the cosmic time $t$, $H=\dot{a}/a$ is the Hubble parameter, and $q=- \ddot{a}/(a H^2)$ is the decelerating parameter. We see that the statefinder parameters are expressed in terms of the third derivative of the scale factor with respect to the cosmic time, contrary to the Hubble parameter and the decelerating parameter, which are expressed in terms of the first and the second time derivative of the scale factor, respectively. It is straightforward to verify that for the $\Lambda$CDM model without radiation the statefinder parameters take constant values, $r=1,s=0$. These parameters may be computed within a certain model, their values can be extracted from future observations \cite{SNAP1,SNAP2}, and the statefinder diagnostic has been applied to several dark energy models \cite{diagnostics1,diagnostics2,diagnostics3,diagnostics4,diagnostics5}.
As we will see later on, $r,s$ can be very different from one model to another even if they predict very similar expansion histories.

Considering that running vacuum (RVM) models offers an interesting framework to study phenomenology
beyond to $\Lambda$CDM model, the main goal of the present work is to analyse 
three models within the running vacuum dynamics: we first consider two vacuum models where dark energy interacts with dark matter \cite{sola7}, and secondly, we investigate a third model where the gravitational coupling is assumed to be a slowly-varying function of the Hubble rate \cite{Fritzsch:2016ewd} and dark energy and dark matter interact as well. The analysis is performed in two respects: On the one hand, by applying the dynamical systems methods, we compute the critical points for each scenario and study their stability. On the other hand, in other to discriminate between the several running vacuum DE models and $\Lambda$CDM, we perform the statefinder diagnostic by means computing the statefinder parameters as a function of the redshift, studying their high and low-redshift limits. Our work is organized as follows: after this introduction, we present the basic equations and analytical solutions for Models I and II in sections 2 and 3, respectively. In the fourth section, upon the dynamical system analysis we compute the corresponding critical points for Models I and II, while in the fifth section we discuss the statefinder parameters for the same models. In
section 6, we present the main results for Model III regarding the dynamical system and statefinder analysis. Finally we summarize our findings and present our conclusions in Section 7. We adopt the mostly positive metric signature, $(-,+,+,+)$, and we work in natural units where $c=\hbar=1$.

%%%%%%%%%%%%%%%%%%%%%%%%%%%%%%%%%%
\section{Theoretical framework}
%%%%%%%%%%%%%%%%%%%%%%%%%%%%%%%%%%

We consider a flat ($k=0$) FLRW Universe
\begin{equation}
ds^2 = -dt^2 + a(t)^2 \delta_{ij} dx^i dx^j,
\end{equation}
and setting $\kappa^2 = 8 \pi G$, with $G$ being the Newton's constant, the scale factor $a(t)$ satisfies the Friedmann equations
\begin{eqnarray}
H^2 & = & \frac{\kappa^2}{3}\sum_{A}\rho_A, \\
\dot{H} & = & - \frac{\kappa^2}{2} \sum_{A}(\rho_A + p_A),
\end{eqnarray}
where $\rho_A$ and $p_A$ denote the energy density and pressure of each individual fluid component, respectively. The equation-of-state parameter for each fluid component $p_A=w_A \rho_A$ takes the values: $w=0$ for baryons and dark matter, $w=1/3$ for radiation and $w=-1$ for DE.

The system of cosmological equations also includes the conservation equations for the non-interacting fluids (baryons, radiation)
\begin{eqnarray}
\dot{\rho}_b + 3 H \rho_b &=&0,\\
\dot{\rho}_r + 4 H \rho_r &=&0,
\end{eqnarray}
as well as for the interacting components (DE and dark matter)
\begin{eqnarray}
\label{rholambda0}
\dot{\rho}_\Lambda & = & -Q, \\
\dot{\rho}_{\textup{dm}} + 3 H \rho_{\textup{dm}} & = & Q.
\label{rhom0}
\end{eqnarray}
Here, $Q$ represents the source term, i.e. the energy exchange between DE and DM. Particularly, in the running vacuum cosmology scenario, the cosmological coupling $\Lambda$ varies as $\Lambda \equiv \Lambda(H^2)$ or $\Lambda \equiv \Lambda(R)$ \cite{Perico:2016kbu}. In such a models, the dynamics of vacuum is due to the energy exchange with some of the fluid components that participate to the evolution of the Universe. 
As running vacuum models (RVM) seem to perform better than the $\Lambda$CDM in some circumstances, in the present work we first consider two scenarios, namely I and II, found e.g. in \cite{sola7} and precisely labelled as ``running
vacuum model" (RVM):
\begin{eqnarray}
%Q_1 & = & \nu H (3 \rho_{m} + 4 \rho_{r}), \\
Q_1 & = & 3 \nu_{\textup{dm}} H \rho_{\textup{dm}}, \\
Q_2 & = & 3 \nu_{\Lambda} H \rho_\Lambda,
\end{eqnarray}
where the dimensionless parameters $\{\nu_i \}$ measure the strength of the interaction term $Q_i$. In the present paper we are interested in studying the late-times cosmology within the interacting vacuum energy scenarios. So, we have neglected the coupling to radiation and baryons because at lower redshifts, $z\ll 2$, the contribution to the total energy density coming from these components is smaller than the dark energy and dark matter components. It is worth to mention that an eventual coupling between radiation and dark energy could have a significant effect on the dynamics of early Universe, see e.g. \cite{copeland}. Particularly, within the interacting vacuum energy, the nucleosynthesis sets strong constraints on the
strength of the coupling, which becomes much smaller than the unity \cite{EspanaBonet:2003vk}.

Following previous works \cite{KPT,ellis,wands,lazkoz,mena} we introduce normalized densities (dimensionless, positive quantities)
\begin{equation}
\Omega_A = \frac{\rho_A}{\rho_{cr}},
\end{equation}
where $\rho_{cr}=3H^2/\kappa^2$ is the critical energy density. On the one hand, the first Friedmann equation is a constraint
\begin{equation}
\Omega_r + \Omega_\Lambda + \Omega_{dm} + \Omega_b = 1,
\end{equation}
or 
\begin{equation}
\Omega_r + \Omega_\Lambda + \Omega_m = 1,
\end{equation}
where $\Omega_m \equiv \Omega_{dm} + \Omega_b$. Because of the constraint, there are either two or three independent normalized densities depending on the interacting model. 
In particular, in scenario I there are three, $\Omega_r, \Omega_\Lambda, \Omega_b$, in contrast, in 
the scenario II there are two, namely $\Omega_r, \Omega_\Lambda$, with the third one being $\Omega_m = 1 - \Omega_r - \Omega_\Lambda$, while the fourth $\Omega_{dm} = 1 - \Omega_r - \Omega_\Lambda - \Omega_b$.
On the other hand, the second Friedmann equation takes the form
\begin{equation}
-\frac{\dot{H}}{H^2} = \frac{3}{2} (1 + w_T),
\end{equation}
where we have defined the total equation-of-state parameter $w_T=p_T/\rho_T$, which is given by
\begin{equation}
w_T = \sum_A w_A \Omega_A = -\Omega_\Lambda + \frac{\Omega_r}{3}.
\end{equation}
Finally, instead of cosmological time $t$ we introduce the number of $e$-folds $N\equiv \ln(a)$, and we define the time derivatives for any quantity A as follows
\begin{eqnarray}
\dot{A} & = & \frac{d A}{d t}, \\
A' & = & \frac{d A}{d N},\\
\dot{A} & = & H A'.
\end{eqnarray}
Using the definitions and the cosmological equations one can obtain first order differential equations for $\Omega_A$ with respect to $N$. The equations for $\Omega_r,\Omega_b$ are the same in all the scenarios since they are non-interacting components
\begin{eqnarray}
\Omega_r' & = & \Omega_r (-1 + \Omega_r - 3 \Omega_\Lambda),  \label{omegar}
\\
\Omega_b' & = & \Omega_b (\Omega_r - 3 \Omega_\Lambda). \label{omegab}
\end{eqnarray}
The equation for $\Omega_\Lambda$ depends on the interaction term $Q$, and therefore there are 3 cases

\begin{widetext}
\begin{equation}
\Omega_\Lambda' =
\left\{
\begin{array}{lcl}
% 3 (\Omega_\Lambda - \nu) \left (1 + \frac{\Omega_r}{3} - \Omega_\Lambda \right), & \mbox{ for } & \text{Model I} 
%\\
%&
%&
%\vspace{0.3cm}
%\\
3 \left[ \Omega_\Lambda  \left (1 + \frac{\Omega_r}{3} - \Omega_\Lambda \right) - \nu_{\textup{dm}} (1 - \Omega_r - \Omega_\Lambda - \Omega_b) \right],  & \mbox{ for } & \text{Model I} \label{omegalambda}
%&
%&
\vspace{0.3cm}
\\
3 \Omega_\Lambda \left (1 + \frac{\Omega_r}{3} - \Omega_\Lambda -\nu_{\Lambda} \right).   & \mbox{ for } & \text{Model II}
\end{array}
\right.
\end{equation}
\end{widetext}
Finally, $q$ and $r$ are computed to be
\begin{eqnarray}
q & = & -1 + \frac{3}{2} \left(1 + \frac{\Omega_r}{3} - \Omega_\Lambda \right),  \\
r & = & -q' + 3 q \left(1 + \frac{\Omega_r}{3} - \Omega_\Lambda \right).
\end{eqnarray}
while $s$ can be computed using its definitions once $q$ and $r$ are known.
Thus, we can compute the statefinder parameters, $\{r,s\}$, as a function of the red-shift  $z \equiv a_0/a - 1$ (with $a_0$ being the present value of the scale factor $a$), after solving the system of differential equations given by Eqs. \eqref{omegar}-\eqref{omegalambda} in three different models for the dimensionless densities $\Omega_{A}$.
Although a numerical integration of the cosmological equations to obtain $\{r,s\}$ is possible, in the following we will obtain exact analytical expressions, see next section.
% 
%We summarise our result in Fig. \eqref{fig:1} left, where we observe that $r$ starts from a constant value, and after that, it grows when redshift grows too. Also, we notice that the parameter $s$ begins from a fixed value, and it evolves by decreasing when $z$ increases.

%%%%%%%%%%%%%%%%%%%%%%%%%%%%%%%%%%%%
\section{Analytical Solutions}
%%%%%%%%%%%%%%%%%%%%%%%%%%%%%%%%%%%%

The system of coupled equations may be directly integrated to obtain concrete expressions for the energy densities in terms of the scale factor, as was done e.g. in Ref.\cite{sola7}. Although these solutions were previously reported, neither the statefinder diagnostic nor the phase space were analysed. This is precisely the goal of the present article, filling thus a gap in the literature. in this paper we want to complete analysis by including the statefinder diagnostic showing, in figures, how the set $\{r,s\}$ evolves for different values of redshift, as well as the phase space of the above parameters.
We start by considering the corresponding dark matter density $\rho_{\text{dm}}$ and dark energy density $\rho_{\Lambda}$ respect to the scale factor for each model, i.e.:
\begin{widetext}
\begin{equation}
\rho_{\text{dm}} =
\left\{
\begin{array}{lcl}
%\rho_{\text{dm}}^0 a^{-3(1-\nu)} + \rho_{b}^0 \left(a^{-3(1-\nu)} - a^{-3}\right) - \frac{4 \nu}{1+3 \nu}%\rho_{r}^0 \left( a^{-4} - a^{-3(1-\nu)}  \right),
%& \mbox{ for } & \text{Model I} 
%\vspace{0.3cm}
%\\
%&
%&
%\\
\rho_{\text{dm}}^0 a^{-3(1-\nu_{\text{dm}})},  & \mbox{ for } & \text{Model I} \label{omegalambda}
%&
%&
\vspace{0.3cm}
\\
\rho_{\text{dm}}^0 a^{-3} + \frac{\nu_{\Lambda}}{1-\nu_{\Lambda}}\rho_{\Lambda}^0 \left( a^{-3 \nu_{\Lambda}} - a^{-3}\right). & \mbox{ for } & \text{Model II}
\end{array}
\right.
\end{equation}
\begin{equation}
\rho_{\Lambda} =
\left\{
\begin{array}{lcl}
%\rho_{\Lambda}^0 + \frac{\nu}{1-\nu}\rho_m^0 \left( a^{-3(1-\nu)}-1\right)+
%\frac{\nu}{1-\nu}\rho_r^0 \left( \frac{1-\nu}{1+3\nu} a^{-4} + \frac{4\nu}{1+3\nu} a^{-3(1-\nu)} - %1\right),
%& \mbox{ for } & \text{Model I} 
%\\
%&
%&
%&\vspace{0.3cm}
%&\\
\rho_{\Lambda}^0 + \frac{\nu_{\text{dm}}}{1-\nu_{\text{dm}}}\rho_{\text{dm}}^0
\left( a^{-3(1-\nu_{\text{dm}})}-1\right),
& \mbox{ for } & \text{Model I} \label{omegalambda}
%&
%&
\vspace{0.3cm}
\\
\rho_{\Lambda}^0 a^{-3 \nu_{\Lambda}}. & \mbox{ for } & \text{Model II}
\end{array}
\right.
\end{equation}
\end{widetext}
Thus, for each particular model the above profile densities give the evolution of dark matter and dark energy respectively. It is important to point out that the models analysed here boil down to the $\Lambda$CDM model when $\nu_i \rightarrow 0$.
Finally, for convenience, we introduce the dimensionless Hubble rate $E(z) \equiv H(z)/H_0$, where $H_0 = 100 h$ (km sec$^{-1}$)/Mpc) is the Hubble constant. Accordingly, the parameters $\{q,r,s\}$ are computed as follows
\begin{align}
q(z) & = -1 + (1 + z) \frac{E_z(z)}{E(z)},
\label{qz}
\\
r(z) &= q(z)(1 + 2q(z)) + (1+z)q_z(z),
\label{rz}
\end{align}
and $s(z)$ is given by \eqref{rsz}, where $X_z \equiv dX/dz$ for any quantity $X$. Using the expressions for the energy densities shown before, one can obtain exact analytical expressions for all quantities of interest versus red-shift, $E(z),q(z),r(z),s(z)$, see section V.

%%%%%%%%%%%%%%%%%%%%%%%%%%%%%%%%%%%%
\section{Dynamical systems methods}
%%%%%%%%%%%%%%%%%%%%%%%%%%%%%%%%%%%%

%\begin{table*}[ht]
%\centering
%\begin{tabular}{ccccc}
%  \hline
%Fixed point &  $(\Omega_r,\Omega_\Lambda)$ & Eigenvalues & $w_T$ & $q$  \\
%\hline
%I.a & $(0,1)$ & $-4$, $-3 (1-\nu)$ & $-1$ & $-1$  \\
%I.b & $(0, \nu)$ & $-1-3 \nu$, $3 (1-\nu)$ & $-\nu$ & $(1-3 \nu)/2$  \\
%I.c & $(1+3 \nu, \nu)$ & $4$, $1+3 \nu$ & $1/3$ & $1$ \\
%\hline
%\end{tabular}
%\caption{Fixed points of model I}
%	\label{table:firstset}
%\end{table*}

%\begin{table*}[ht]
%\centering
%

%\begin{tabular}{cccc}
%  \hline
%Fixed point & Existence & Acceleration & Nature \\
%\hline
%(0,1) & $\vee \: \nu$ & $\vee \: \nu$ & A ($\nu < 1$), S ($\nu > 1$) \\
%($0,\nu$) & $0 < \nu < 1$ & $\nu > 1/3$ & S  \\ 
%\hline
%\end{tabular}
%  \caption{Nature of fixed points of model I}
%\label{table:secondset}
%\end{table*}

We briefly review the stability analysis based on the nature of the fixed points (FPs), see e.g. \cite{KPT,ellis,wands,lazkoz,mena}. Suppose that for a dynamical system with a two-dimensional phase space $(x,y)$, its time evolution is determined by the following system of coupled first order differential equations
\begin{eqnarray}
\frac{d x}{d t} & = & F(x(t),y(t)),  \\
\frac{d y}{d t} & = & G(x(t),y(t)).
\end{eqnarray}
First, the fixed point(s) is (are) computed setting $dx/dt=0=dy/dt$, and one has to solve the system of two algebraic equations $F(x_0,y_0)=0=G(x_0,y_0)$. Then, to determine the nature of the fixed point(s) we linearise the equations around that point, $x(t)=x_0 + \delta x, y(t)=y_0 + \delta y$
ignoring higher order terms. One obtains a system of two coupled linear equations of the form
\begin{equation}
\dot{X} = A X,
\end{equation}
where the column $X$ contains the two functions $\delta x(t), \delta y(t)$, while $A$ is a two-dimensional matrix, the elements of which are given by
\begin{eqnarray}
A_{11} & = & F_x(x_0,y_0),  \\
A_{12} & = & F_y(x_0,y_0),  \\
A_{21} & = & G_x(x_0,y_0),  \\
A_{22} & = & G_y(x_0,y_0).  
\end{eqnarray}
Finally, we compute the eigenvalues $\lambda_1,\lambda_2$ of $A$, the sign of which determines the nature of the fixed point(s). In particular, the critical point is stable (A) when both eigenvalues are negative, unstable (R) when both eigenvalues are positive, and a saddle point (S) if the eigenvalues are of opposite sign. Furthermore, if $q(x_0,y_0) < 0, w_T(x_0,y_0) < -1/3$, the fixed point at hand corresponds to acceleration. The procedure may be easily generalized in a straightforward manner for a three-dimensional phase-space.

The fixed points and their nature (stability conditions) for all two models considered in this work are shown in the Tables 
%\ref{table:firstset},\ref{table:secondset},
\ref{table:thirdset},\ref{table:fourthset},\ref{table:fifthset}, and \ref{table:sixthset}.

%The third point of Model I is not acceptable, since either $\nu$ or $1+3 \nu$ lies outside the range (0,1) irrespectively of the sign of $\nu$.

%\subsection{Model I}

\begin{table*}[ht]
\centering

\begin{tabular}{ccccc}
  \hline
Fixed point & $(\Omega_r,\Omega_\Lambda,\Omega_b)$ & Eigenvalues & $w_T$ & $q$  \\
\hline
I.a & $(0,0,1)$ & $3$, $-1$,  $3 \nu_{\textup{dm}}$ & $0$ & $1/2$  \\
I.b & $(1,0,0)$ & $4$, $1$, $1+3 \nu_{\textup{dm}}$ & $1/3$ & $1$  \\
I.c & $(0,1,0)$ & $-4$, $-3$, $-3(1-\nu_{\textup{dm}})$ & $-1$ & $-1$ \\
I.d & $(0,\nu_{\textup{dm}},0)$ & $-1-3 \nu_{\textup{dm}}$, $3 (1-\nu_{\textup{dm}})$, $-3 \nu_{\textup{dm}}$ & $-\nu_{\textup{dm}}$ & $(1-3 \nu_{\textup{dm}})/2$  \\
\hline
\end{tabular}
  \caption{Fixed points of model I}
\label{table:thirdset}
\end{table*}

\begin{table*}[ht]
\centering
\begin{tabular}{cccc}
  \hline
Fixed point & Existence & Acceleration & Nature \\
\hline
(0,0,1) & $\vee \: \nu_{\textup{dm}}$ & No & S \\
(1,0,0) & $\vee \: \nu_{\textup{dm}}$ & No & S ($\nu_{\textup{dm}} < -1/3$), R ($\nu_{\textup{dm}} > -1/3$) \\ 
(0,1,0) & $\vee \: \nu_{\textup{dm}}$ & $\vee \: \nu_{\textup{dm}}$  & S ($\nu_{\textup{dm}} > 1$), A ($\nu_{\textup{dm}} < 1$) \\
(0, $\nu_{\textup{dm}}$, 0) & $0 < \nu_{\textup{dm}} < 1$ & $\nu_{\textup{dm}} > 1/3$ & S \\
\hline
\end{tabular}
  \caption{Nature of fixed points of model I}
\label{table:fourthset}
\end{table*}

\subsection{Model I}

\begin{table*}[ht]
\centering
\begin{tabular}{ccccc}
  \hline
Fixed point & $(\Omega_r,\Omega_\Lambda)$ & Eigenvalues & $w_T$ & $q$  \\
\hline
II.a & $(0,0)$ & $-1$, $3 (1-\nu_{\Lambda})$ & $0$ & $1/2$  \\
II.b & $(1,0)$ & $1$, $4-3 \nu_{\Lambda}$ & $1/3$ & $1$  \\
II.c & $(0, 1-\nu_{\Lambda})$ & $-4+3 \nu_{\Lambda}$, $-3 (1-\nu_{\Lambda})$ & $-1+\nu_{\Lambda}$ & $-1 + 3 \nu_{\Lambda}/2$ \\
\hline
\end{tabular}
  \caption{Fixed points of model II}
\label{table:fifthset}
\end{table*}

\begin{table*}[ht]
\centering
\begin{tabular}{cccc}
  \hline
Fixed point & Existence & Acceleration & Nature \\
\hline
(0,0) & $\vee \: \nu_{\Lambda}$ & No & S ($\nu_{\Lambda} < 1$), A ($\nu_{\Lambda} > 1$) \\
(1,0) & $\vee \: \nu_{\Lambda}$ & No & R ($\nu_{\Lambda} < 4/3$), S ($\nu_{\Lambda} > 4/3$)  \\ 
(0, $1-\nu_{\Lambda}$) & $0 < \nu_{\Lambda} < 1$ & $\nu_{\Lambda} < 2/3$ & A  \\
\hline
\end{tabular}
  \caption{Nature of fixed points of model II}
\label{table:sixthset}
\end{table*}

In this case we obtain four critical points, which are shown in Tables \ref{table:thirdset} and \ref{table:fourthset}. Point I.a is a matter dominated solution representing ordinary baryonic matter, such that $\Omega_{b}=1$, and with $w_{T}=0$. The eigenvalues for this critical point are 
\be 
\mu_{1}=3, \:\:\: \mu_{2}=-1,\:\:\: \mu_3=3 \nu_{\text{dm}},
\ee and therefore it is always a saddle point. This fixed point is not physical because it represents an era dominated by baryons. It is well known that cold dark matter constitutes the dominant component during the matter-dominated era at redshift $1 \lesssim z\lesssim 10^3$, and thus during this epoch it provides the main contribution for structure formation in the universe.

Point I.b corresponds to a radiation dominated solution, $\Omega_{r}=1$, for which one has that $w_{T}=1/3$ and therefore there is not acceleration. For this fixed point we find the eigenvalues 
\be 
\mu_{1}=4,\:\:\: \mu_{2}=1, \:\:\: \mu_{3}=1+3 \nu_{\text{dm}},
\ee which means that it is always an unstable FP for $\nu_{\text{dm}}>0$. 

On the other hand, point I.c is a de Sitter-dominated solution for which $\Omega_{\Lambda}=1$ and $w_{DE}=w_{T}=-1$. So, this solution presents accelerated expansion for all values of $\nu_{\text{dm}}$. In this case, we find the eigenvalues 
\be
\mu_{1}=-4,\:\:\: \mu_{2}=-3,\:\:\: \mu_{3}=-3(1-\nu_{\text{dm}}).
\ee Clearly, for $\nu_{\text{dm}}<1$, point I.c is a stable node and therefore an attractor. 

The last solution for this model is the fixed point I.d which is a scaling solution with $\Omega_{\Lambda}=\nu_{\text{dm}}$, as the physical requirement implies $0<\nu_{\text{dm}}<1$. Also, this solution is characterized by $w_{T}=-\nu_{\text{dm}}$, with the decelerating and accelerating regimes satisfying $0<\nu_{\text{dm}}<1/3$ and $\nu_{\text{dm}}>1/3$, respectively. Point I.d behaves as a dark matter solution in the limit $\nu_{\text{dm}}\ll 1$, with a small contribution of dark energy proportional to $\nu_{\text{dm}}$, during the matter dominated epoch, and thus suppressing the growth of matter perturbations. Stability analysis leads us to the eigenvalues 
\be
\mu_{1}=-1-3\nu_{\text{dm}},\:\:\: \mu_{2}=3\left(1-\nu_{\text{dm}}\right),\:\:\: \mu_{3}=-3 \nu_{\text{dm}}.
\ee Since we require $0<\nu_{\text{dm}}<1$, the point I.d is always a saddle point. 

The critical point I.d is a dark matter dominated solution for the model I, with a small contribution from dark energy density given by $\Omega_{\Lambda}=\nu_{dm} \ll 1$ and  total equation of state $w_{T}=-\nu_{dm}\approx 0$. Although this fixed point can provide accelerated expansion for $\nu_{dm}>1/3$, the observational constraints on $\nu_{dm}\ll 1$ \cite{EspanaBonet:2003vk} do not allow that this happens. So, the thermal history of the Universe is successfully reproduced for model I provided it satisfies the restriction $\nu_{\text{dm}}\ll 1$. 

It is also important to note that in the present framework of dynamical systems the negative values for $\nu_{dm}$ are excluded since this would imply a negative energy density $\Omega_{\Lambda}=\nu_{dm}<0$. So, as several authors usually do, we have given preference to maintain the physical condition $\rho\geq 0$ in agreement with the weak energy condition (WEC) \cite{copeland}.

\subsection{Model II}

\begin{figure*}[ht!]
\centering
%\includegraphics[width=0.32\textwidth]{Figq03.pdf}   \
%\includegraphics[width=0.32\textwidth]{Figq007.pdf}  \
%\includegraphics[width=0.32\textwidth]{Figq0007.pdf} \
%
%\medskip
%
%\includegraphics[width=0.32\textwidth]{Figr03.pdf} \
%\includegraphics[width=0.32\textwidth]{Figr007.pdf} \
%\includegraphics[width=0.32\textwidth]{Figr0007.pdf}
%
%\medskip
%
%\includegraphics[width=0.32\textwidth]{Figs03.pdf} \
%\includegraphics[width=0.32\textwidth]{Figs007.pdf} \
%\includegraphics[width=0.32\textwidth]{Figs0007.pdf}
%
%\medskip
%
\includegraphics[width=0.48\textwidth]{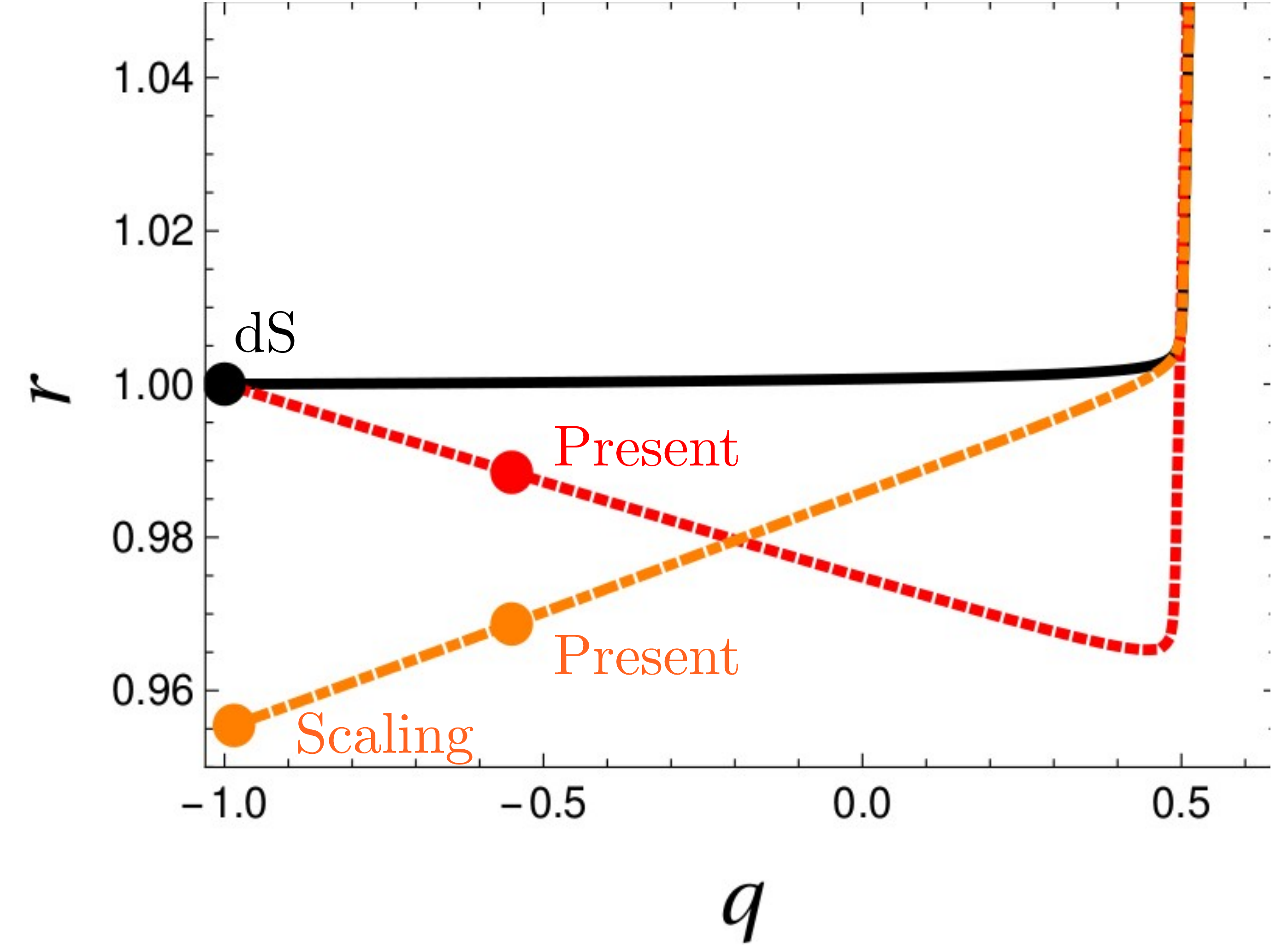} \
\includegraphics[width=0.48\textwidth]{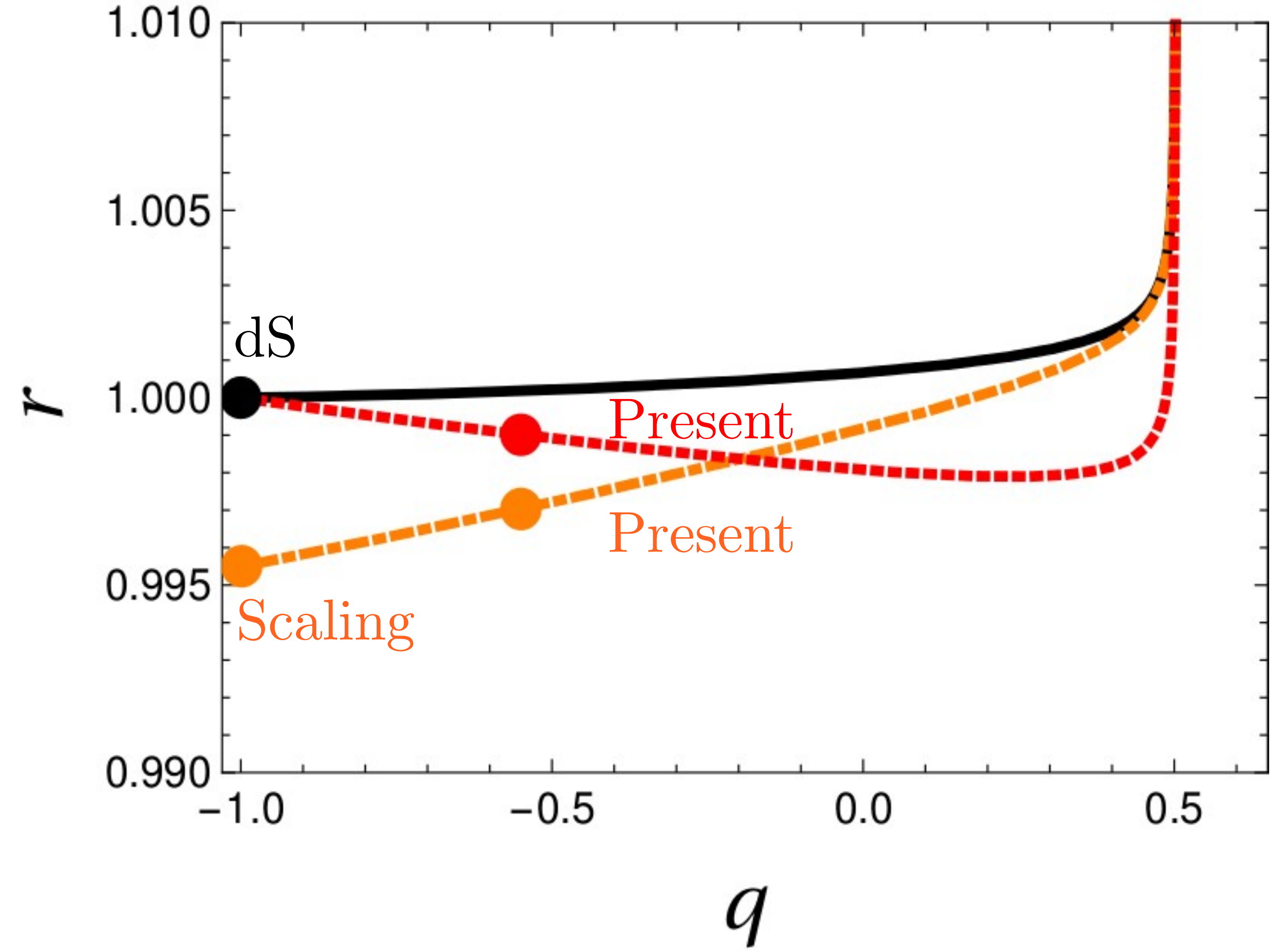} \

\medskip

\includegraphics[width=0.48\textwidth]{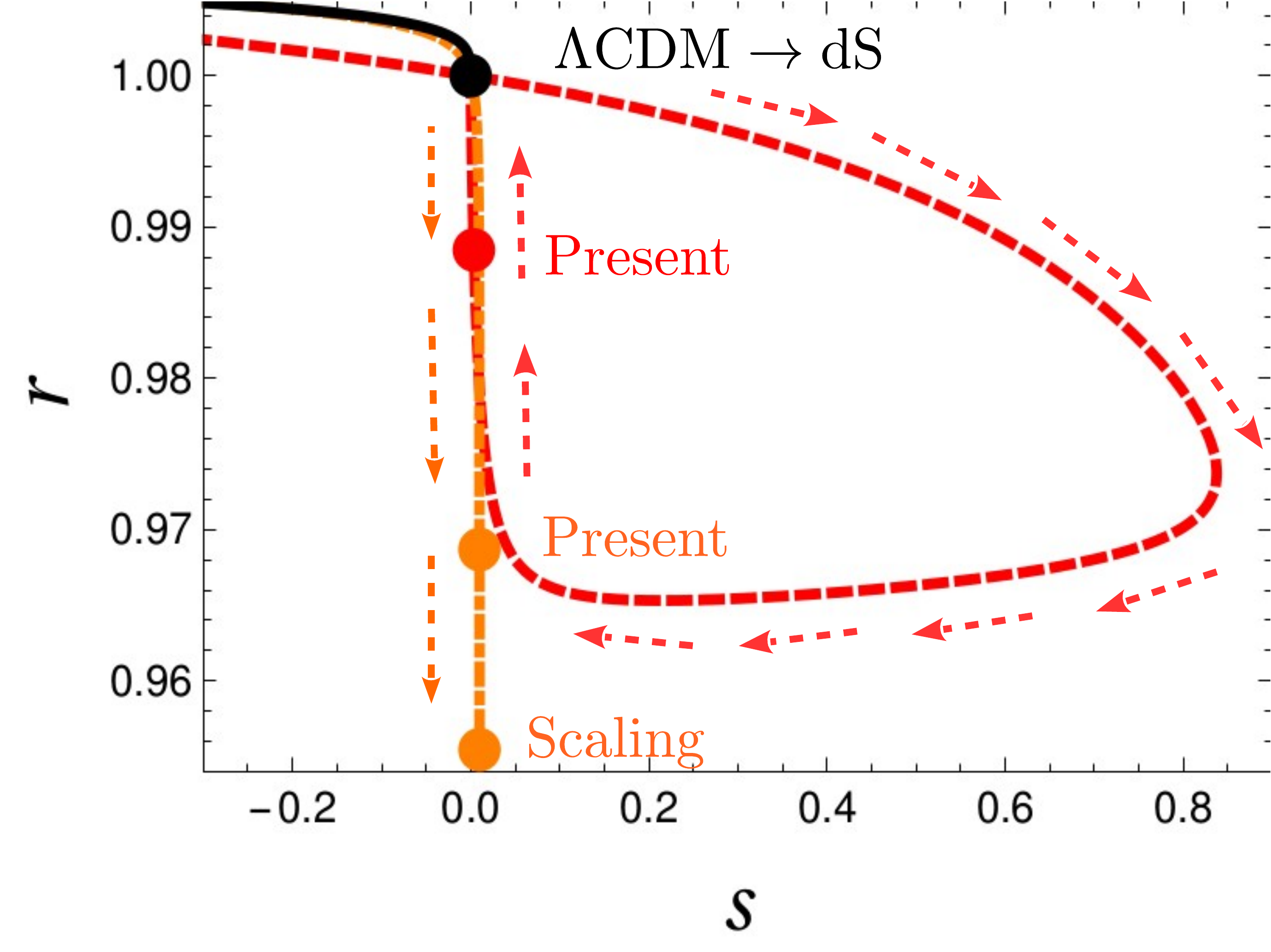} \
\includegraphics[width=0.48\textwidth]{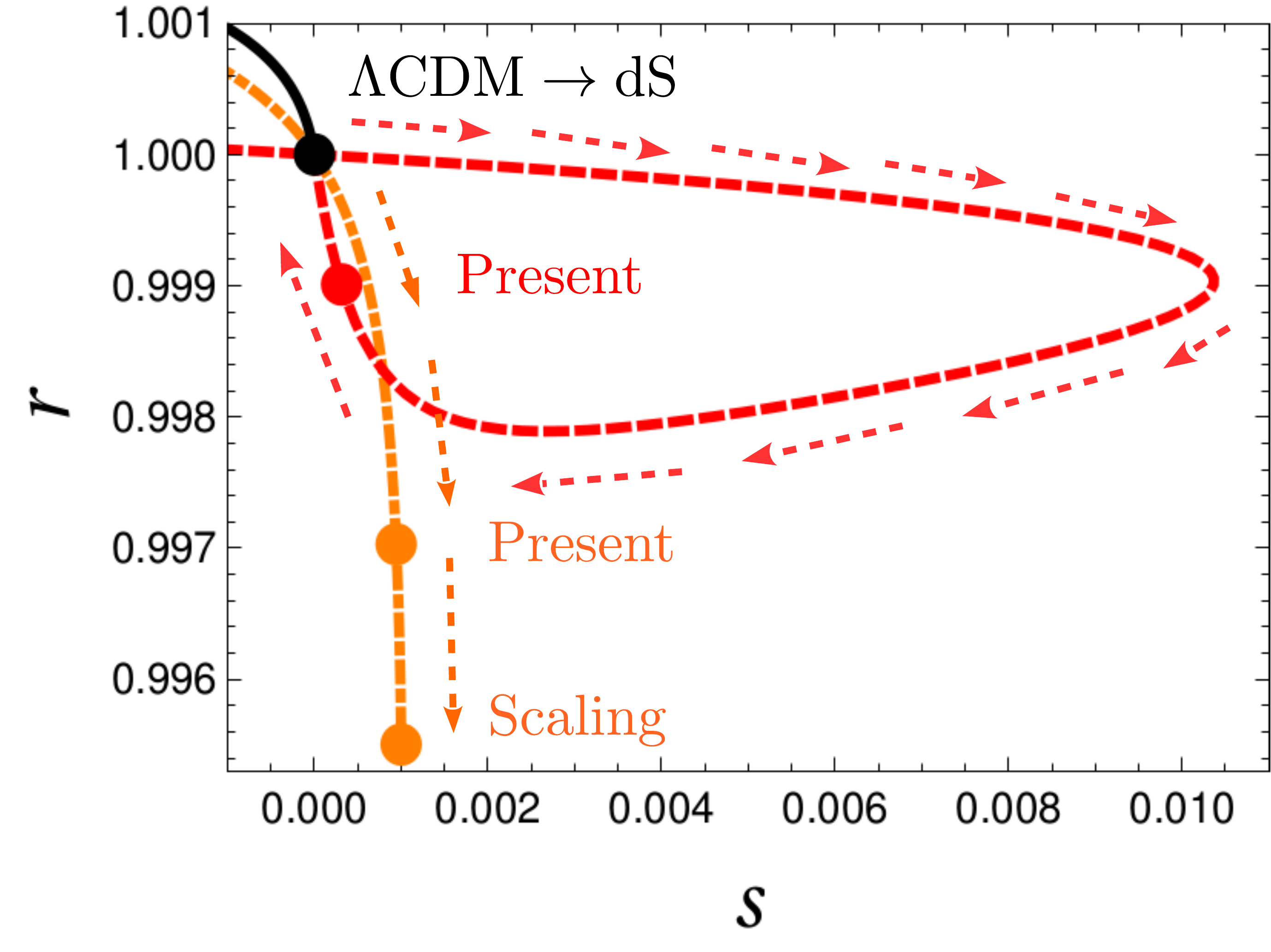} \

\caption{
 The figures show the 
%evolution of the parameters $\{q,r,s\}$ versus red-shift for the three models analysed here and for $\Lambda$CDM. Besides, we plot the 
parameter space $q-r$ and $s-r$ for the two first models. To show the impact of the parameter $\nu_{i}=\{\nu_{\text{dm}},\nu_{\Lambda}\}$, we compute the functions for two different values of it, i.e.: 
ii) $\nu_{i}=0.01$ (first column), and finally
iii) $\nu_{i}=0.001$ (second column).
In addition, the color code is as follow: i) solid black line correspond to $\Lambda$CDM, ii) dashed red line correspond to first model and finally, iii) dot-dashed orange line correspond to second model. 
}
\label{fig:3}
\end{figure*}

Model II, has three critical points which are shown in Tables \ref{table:fifthset} and \ref{table:sixthset}. Point II.a is a dark matter dominated solution for the which $\Omega_{m}=1$, and $w_{T}=0$. The eigenvalues associated with this critical point are 
\be 
\mu_{1}=-1\:\:\:\: \mu_{2}=3\left(1-\nu_{\Lambda}\right),
\ee and hence, one can see that it is always a saddle point for the value $0<\nu_{\Lambda}<1$. 

On the other hand, point II.b is a solution for which radiation component is the dominant one, being that $\Omega_{r}=1$ and $w_{T}=1/3$. From stability analysis we obtain the eigenvalues 
\be 
\mu_{1}=1 \:\:\:\: \mu_{2}=4-3 \nu_{\Lambda},
\ee
 which means that it is an unstable node for $\nu_{\Lambda}<4/3$ and a saddle in the opposite case $\nu_{\Lambda}>4/3$. 

Finally, the point II.c is a scaling solution with $\Omega_{\Lambda}=1-\nu_{\Lambda}$ that, interestingly, can also be an attractor with $w_{T}=-1+\nu_{\Lambda}$, allowing to alleviate the so-called cosmological coincidence problem \cite{copeland}. The physical requirement $0<\Omega_{\Lambda}<1$ implies $0<\nu_{\Lambda}<1$, and from the constraint $w_{T}<-1/3$, the accelerated expansion occurs for the value $\nu_{\Lambda}<2/3$. For this critical point one finds the eigenvalues 
\be 
\mu_{1}=-4+3 \nu_{\Lambda},\:\:\:\: \mu_{2}=-3 \left(1-\nu_{\Lambda}\right),
\ee 
So, point II.c is a stable FP for $\nu_{\Lambda}<4/3$ and a saddle point for $4/3<\nu_{\Lambda}<1$. Like model I, the present model II is also physically viable to successfully reproduce the thermal history of the Universe from the radiation dominated era, going through the standard matter dominated era, to late times when the dark energy component dominates the total energy density and pressure of the Universe.

Considering the observational constraints as, for instance, those found in  \cite{Kumar:2019wfs,DiValentino:2019ffd} (and references therein) the value of the strength of the coupling is quite smaller that unity, such that $\nu_{\Lambda}\lesssim 0.01$, and this is in agreement with our theoretical bounds. So, the results obtained from the dynamical analysis in this section should be supplemented by the observational bounds. In fact, although the scaling solution for model II  allows us to adjust $\Omega^{0}_{\Lambda}\simeq 0.73$ and $\Omega^{0}_{m}=0.27$ for $\nu_{\Lambda}=0.3$, actually this solution cannot reached at $z=0$, but only asymptotically to reproduce the whole thermal history of the universe. In other words, in order to obtain the physical trajectory in the phase space consistently with observational data, this scaling solution only can be reached at the future, in such a way that we need to choice smaller values of $\nu_{\Lambda}$ ($\lesssim 10^{-2})$, allowing to have $\Omega^{0}_{\Lambda}\simeq 0.73$ at $z=0$, and  asymptotically $\Omega_{\Lambda}=1-\nu_{\Lambda}\simeq 1$ with $w_{T}=1-\nu_{\Lambda}\simeq -1$ for $z\rightarrow-1$.

\section{Statefinder analysis}

In the present analysis we have calculated for all three models I, and II, the analytical expression of the Hubble parameter $E(z)$ as explicit function of the red-shift $z$, and then we have obtained the corresponding functions $q(z)$, $r(z)$ and $s(z)$.  It is important to observe that in the $s-r$ plane, the flat $\Lambda$CDM scenario correspond to the point $(0,1)$, while that in the $q-r$ plane the point $(-1,1)$ is the asymptotic de Sitter solution \cite{Granda:2013gka}.
Since we are only focused on the evolution at late times, particularly in the transition from matter dominated era to the present time, we neglect the radiation component in the computation of $E(z)$ and $\{q,r,s\}$. However, in order to produce the plots shown in Fig.(\ref{fig:3}) we take into account the contribution coming from radiation.

For Model I we find
\bea
&& E(z)=\Bigg[\frac{\Omega^{0}_{\text{dm}}\left[(z+1)^{3\left(1-\nu_{\text{dm}}\right)}-1\right]}{1-\nu_{\text{dm}}}+\nonumber\\
&& \Omega^{0}_{b} \left((z+1)^3-1\right)+1\Bigg]^{\frac{1}{2}},
\eea and thus, for the state-finder parameters we have

\bea
&& q(z)=-1+\Bigg[\frac{3(\nu_{\text{dm}}-1) \left(1+\frac{\Omega^{0}_{b}(z+1)^{3 \nu_{\text{dm}}}}{\Omega^{0}_{\text{dm}}}\right)}{2}\Bigg]\times\nonumber\\
&& \Bigg[\frac{1+\frac{(\nu_{\text{dm}}-1)\Big[\Omega^{0}_{b} z (z (z+3)+3)+1\Big]}{\Omega^{0}_{\text{dm}}}}{(z+1)^{3 (1-\nu_{\text{dm}})}}-1\Bigg]^{-1},\\
&& r(z)=1+\Bigg[\frac{9 (1-\nu_{\text{dm}})\nu_{\text{dm}}}{2}\Bigg]\times\nonumber\\
&& \Bigg[\frac{1+\frac{(\nu_{\text{dm}}-1) (\Omega^{0}_{b} z (z (z+3)+3)+1)}{\Omega^{0}_{\text{dm}}}}{(z+1)^{3 (1-\nu_{\text{dm}})}}-1\Bigg]^{-1},\\
&& s(z)=\frac{(1-\nu_{\text{dm}}) \nu_{\text{dm}}}{\nu_{\text{dm}}+\frac{(\nu_{\text{dm}}-1) (\Omega^{0}_{b}-1)-{\Omega^{0}_{\text{dm}}}}{{\Omega^{0}_{\text{dm}}} (z+1)^{3 (1-\nu_{\text{dm}})}}}.
\eea
Now, the Hubble rate at $z\gg1$ yields
\bea
H\sim \frac{2}{3 t},
\label{H_ModelII_Matter_Epoch}
\eea while the parameters $q$, $r$ and $s$ behaves as
\bea
&& q= \frac{1}{2},\\
&& r= 1,\\
&& s= 1-\nu_{\text{dm}}.
\eea Therefore, in this case we recover the standard matter-dominated era with $0<s<1$. 

In the limit $z\rightarrow -1$, the model predicts a de Sitter solution with
\be
H=H_{0}\sqrt{1-\Omega^{0}_{b}+\frac{\Omega^{0}_{dm}}{1-\nu_{\text{dm}}}},
\ee such that $q=-1$, $r=1$ and $s=0$. Hence, at the present, $z=0$, we obtain the values 
\bea
&& q_{0}=-1+\frac{3}{2}\left(\Omega^{0}_{b}+\Omega^{0}_{\text{dm}}\right)\approx -0.55,\\
&& r_{0}=1-\frac{9 \nu_{\text{dm}}\Omega^{0}_{\text{dm}}}{2},\\
&& s_{0}=\frac{\nu_{2}\Omega^{0}_{\text{dm}}}{1-\Omega^{0}_{b}-\Omega^{0}_{\text{dm}}}.
\eea In calculating some numerical values we take $\nu_{\text{dm}}=\{0.01, 0.001\}$ for which we get $q_{0}\approx -0.55 $, $r_{0}\approx \{0.9883,0.99883 \}$, and $s_{0}=\{0.003714, 0.000314\}$.

Finally, in the case of Model II one finds
\bea
&& E(z)=\left(1-\nu_{\Lambda}\right)^{-\frac{1}{2}} \Big[(1-\Omega^{0}_{b}-\Omega^{0}_{\text{dm}}) (z+1)^{3 \nu_{\Lambda}}+\nonumber\\
&& (z+1)^3 (\Omega^{0}_{b}+\Omega^{0}_{\text{dm}}-\nu_{\Lambda})\Big]^{1/2},
\eea
 and we also we obtain the expressions
\bea
&& q(z)=-1+\frac{3}{2}\left[\frac{1-\frac{\nu_{\Lambda} (\Omega^{0}_{b}+\Omega^{0}_{\text{dm}}-1) (z+1)^{3(\nu_{\Lambda}-1)}}{\Omega^{0}_{b}+\Omega^{0}_{\text{dm}}-\nu_{\Lambda}}}{1-\frac{(\Omega^{0}_{b}+\Omega^{0}_{\text{dm}}-1) (z+1)^{3 (\nu_{\Lambda}-1)}}{\Omega^{0}_{b}+\Omega^{0}_{\text{dm}}-\nu_{\Lambda}}}\right],\\
&& r(z)=\frac{1+\frac{(9 (\nu_{\Lambda}-1) \nu_{\Lambda}+2) (\Omega^{0}_{b}+\Omega^{0}_{\text{dm}}-1) (z+1)^{3 (\nu_{\Lambda}-1)}}{2\left(\nu_{\Lambda}-\Omega^{0}_{b}-\Omega^{0}_{\text{dm}}\right)}}{1+\frac{(\Omega^{0}_{b}+\Omega^{0}_{\text{dm}}-1) (z+1)^{3 (\nu_{\Lambda}-1)}}{\nu_{\Lambda}-\Omega^{0}_{b}-\Omega^{0}_{\text{dm}}}},\\
&& s(z)=\nu_{\Lambda}.
\eea

Similarly as in model I, for $z\gg1$, it is recovering the standard matter-dominated era, 
where the Hubble rate satisfies the relation \eqref{H_ModelII_Matter_Epoch}. Also, in this limit we find
\bea
&& q=\frac{1}{2},\\
&& r=1,\\
&& s=\nu_{\Lambda},
\eea with $0<s<1$.

On the other hand, in the limit $z\rightarrow -1$, unlike models I, the model II behaves as a scaling solution with 
\be
H\sim \frac{\beta}{t}, \:\:\:\:\:\: \beta=\frac{2}{3(1-\nu_{\Lambda})},
\ee with a scale factor of the form $a\sim a_{0}\left(t/t_{0}\right)^\beta$. It is straightforward 
to check that in this limit, the statefinder parameters become
\bea
&& q=-1+\frac{3 \nu_{\Lambda}}{2},\\
&& r=1+\frac{9 \nu_{\Lambda}}{2}\left( \nu_{\Lambda}-1\right),\\
&& s=\nu_{\Lambda}.
\eea 
Thus, for $\nu_{\Lambda}=\{ 0.01, 0.001\}$ one gets $q_{0}\approx\{-0.985,-0.999\} $, $r_{0}\approx \{0.9555,0.9955 \}$, and $s_{0}=\{0.01, 0.001\}$.

In Fig.~\ref{fig:3} we have depicted the behaviour of the parameters $q$, $r$ and $s$ as functions of the red-shift $z$, along with the trajectories of evolution in the $q-r$ and $s-r$ planes, for Models I (red-dotted), and II (orange dash-dotted). For a sake of comparison, we have also included the $\Lambda$CDM model (solid black line). Recall that, in order to produce the plots
shown in Fig.~\ref{fig:3}, the contribution from radiation has been taken into account, and we also set the following values for the fractional energy densities: $\Omega_{\text{dm}}^0 = 0.26$, $\Omega_b^0=0.04$, and $\Omega_r^0=9\times 10^{-5}$. Naturally, $\Omega_m^0 \equiv \Omega_{\text{dm}}^0 + \Omega_b^0$ and $\Omega_{\Lambda}^0 = 1 - \Omega_m^0-\Omega_r^0$.

\begin{figure*}[t!]
\centering
%\includegraphics[width=0.49\textwidth]{r_vs_s.pdf} 
%\
\includegraphics[width=0.8\textwidth]%{Phase_Space_v3.pdf}
{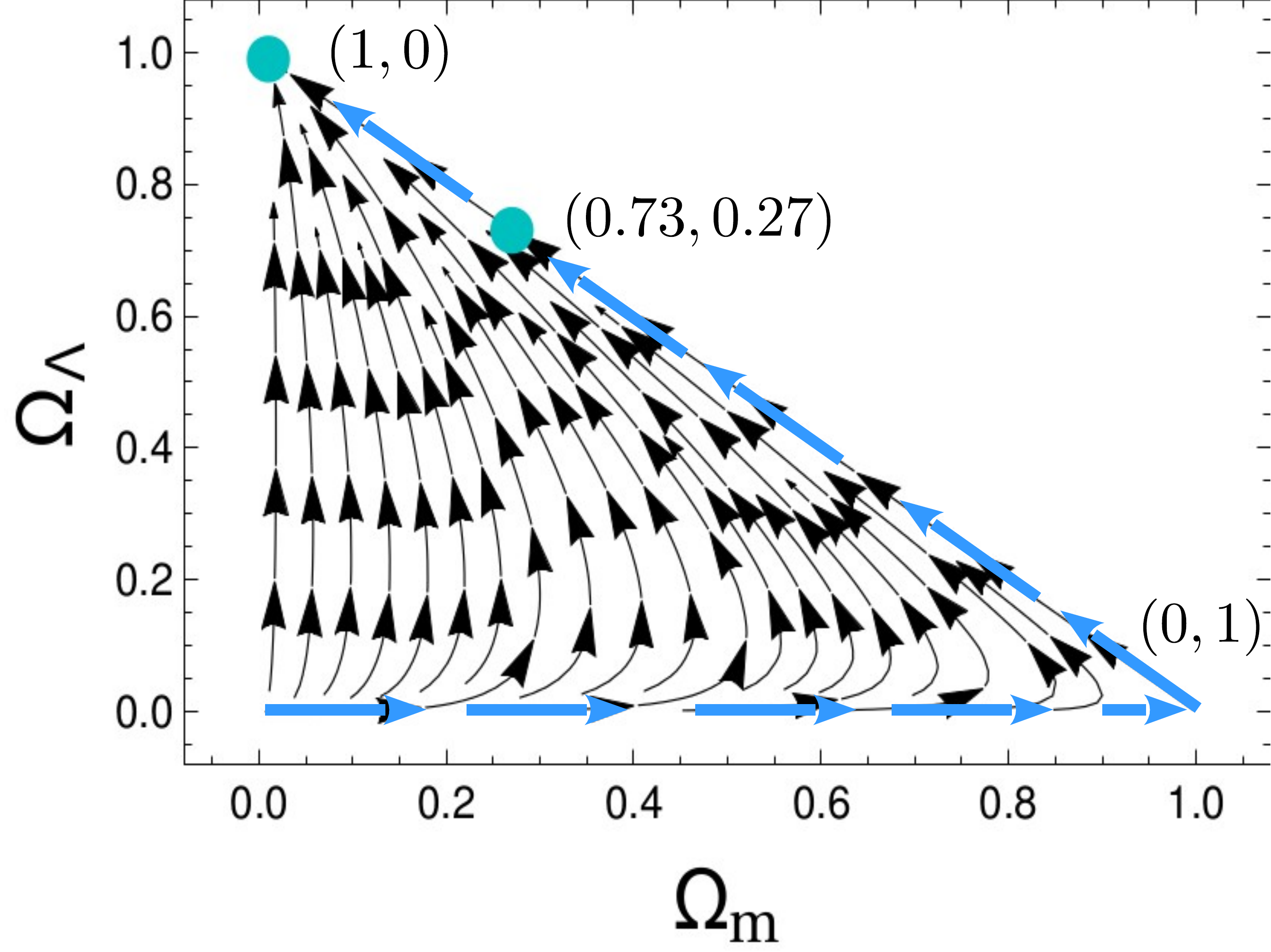}
\caption{
%{\bf Right panel:} 
Two-dimensional phase-space $(\Omega_m-\Omega_\Lambda)$ for Model II and for $\nu_{\Lambda}=0.01$. Different trajectories correspond to different initial conditions. All of them meet at the attractor $\sim (0.0,1.0)$ at late times labelled with a point. Notice that our solution does not reach the de-Sitter solution, but it is undistinguishable from it.
}
\label{fig:5}
\end{figure*}

In these plots the behaviour of the parameters $q(z)$, $r(z)$ and $s(z)$ is in agreement with the analytical results that we have obtained in the limit case of a negligible radiation component for $z\rightarrow -1$. For all the three models I, and II, it is seen that the pair $(s,r)$ starts in the left-hand side of the $\Lambda$CDM fixed point, which is characteristic of the hybrid expansion law (HEL), Chaplygin gas and Galileon models, such that $s<0$ and $r>1$ \cite{Akarsu:2013xha}. Let us notice that this behaviour is very different from what occurs in the case of the quintessence model for which it is observed that the trajectory in the $(s,r)$ plane starts in the region $0<s<1$ and $r<1$.  On the other hand, the trajectory in the $(q,r)$ plane starts in the region bounded by $0<q<1$ and $r>1$, being that in the case of models I, the $\Lambda$CDM line is crossed at some red-shift in the past to then evolve towards the de Sitter fixed point $(q=-1,r=1)$ at the future. For model II, the behaviour becomes different because in this case the asymptotic fixed point is a scaling solution such that for $\nu_{\Lambda}=0.01$, the benchmark values of the fractional energy densities are $\Omega_\Lambda\simeq 1$ and $\Omega_m\simeq 0$, but at $z=0$, we have $\Omega_\Lambda\simeq 0.7 $ and $\Omega_m\simeq 0.3$, as it has been depicted in Fig.~\ref{fig:5}.

%%%%%%%%%%%%%%%%%%%%%%%%%%%%%%%%%%%%%%%%%%%%%%%%%%%%%%%%
\section{Variable gravitational coupling}
%%%%%%%%%%%%%%%%%%%%%%%%%%%%%%%%%%%%%%%%%%%%%%%%%%%%%%%%

In this section we generalize our previous results by allowing a variable gravitational constant within the framework of interacting RVM's.

The generalized Friedmann equations are given by
\bea
\label{H00_G_Var}
&& 3 H^2=8 \pi G\left[\rho_{m}+\rho_{\Lambda}\right],\\
&& \dot{H}= -4 \pi  G\rho_{m},
\label{Hii_G_Var}
\eea where $G(t)$ is a function of the cosmic time $t$, and $\rho_m$ denotes the non-relativistic matter energy density, including both baryons and dark matter. The conservation law for this model can be written as
\be
\dot{\rho}_{m}+\dot{\rho}_{\Lambda}+3 H \rho_{m}+\frac{\dot{G}}{G}\left(\rho_{m}+\rho_{\Lambda}\right)=0.
\ee Following our analysis for interacting RVM's, this equation can also be splitted into a set of two separate evolution equations for the energy densities $\rho_m$ and $\rho_{\Lambda}$, according to Ref. \cite{Fritzsch:2016ewd}, as follows
\bea
&& \dot{\rho}_{m}+3 H \rho_{m}=Q,\\
\label{CEq_rho_m}
&& \dot{\rho}_{\Lambda}=-Q-\frac{\dot{G}}{G}\left(\rho_{\Lambda}+\rho_{m}\right).
\eea Let us note that the two above equations are reduced to the Eqs. \eqref{rholambda0}, \eqref{rhom0}, in the case when the gravitational coupling becomes constant. On the other hand, in order to compare
this model with current observational data regarding dark energy, we rewrite the above Friedmann equations in the standard form 
\bea
\label{Eff_FR_Eq00}
&& 3 H^2=8 \pi G_{0}\left[\rho_{m}+\rho_{de}\right],\\
&& \dot{H}= -4 \pi  G_{0}\left[\rho_{m}+\rho_{de}+p_{de}\right],
\label{Eff_FR_Eqii}
\eea with $G_{0}$ being is the constant gravitational coupling. In these equations we have introduced the effective energy density and pressure density of dark energy, defined as
\bea
&& \rho_{de}=-\rho_{m}+\frac{G}{G_{0}}\left(\rho_{m}+\rho_{\Lambda}\right),\\
&& p_{de}=-\frac{G}{G_{0}}\rho_{\Lambda},
\eea respectively. This effective dark energy density includes the effect of both vacuum energy density $\rho_{\Lambda}(t)$ and the dynamically changing gravitational coupling $G(t)$. It is straightforward to show that it satisfies the evolution equation 
\bea
\dot{\rho}_{de}+3 H (\rho_{de}+p_{de})=-Q.
\label{CEq_rho_DE}
\eea Therefore, it is easy to see that the two evolution equations \eqref{CEq_rho_m} and \eqref{CEq_rho_DE} together are consistent with the energy conservation law for $\rho_{m}$ and $\rho_{DE}$.

As it is usually done, we introduce the fractional and critical energy densities
\be
\Omega_{de}=\frac{\rho_{DE}}{\rho_{cr}},\:\:\:\: \rho_{cr}=\frac{3 H^2}{8 \pi G_{0}},
\label{Omega_de}
\ee the EOS parameter of dark energy 
\be
w_{de}=\frac{p_{de}}{\rho_{de}}=-\frac{G}{G_{0}}+\left(-1+\frac{G}{G_{0}}\right)\frac{1}{\Omega_{DE}},
\label{wde_G_Vary}
\ee and  the total EOS parameter 
\be
w_{T}=\frac{p_{T}}{\rho_{T}}=-1+(1-\Omega_{DE})\frac{G}{G_{0}}. 
\label{wT_G_Vary}
\ee With this, the condition for having accelerated expansion
becomes $w_{eff}<-1/3$, or equivalently $q<0$.

\subsection{An example for variable $G$}

In order to obtain concrete results, we assume the following phenomenological ansatz
for the gravitational coupling depending on the Hubble rate, according to \cite{Fritzsch:2016ewd}
\be
G(X)=\frac{G_{0}}{1+\nu_{G} \ln X},
\label{G_H}
\ee where $X\equiv E^2=\left(H/H_{0}\right)^2$. Also, in order to extend our previous analysis for interacting RVM's, we consider the coupling between DE and DM to be $Q=3 \nu_{m} H \rho_{m}$ (Model II). For this coupling function $Q$ and ansatz \eqref{G_H}, the equations \eqref{CEq_rho_m} and \eqref{CEq_rho_DE} take the form
\bea
&& \frac{d\rho_{m}}{dX}= (1-\nu_{m}) \rho^{0}_{cr} (\nu_{G} \ln X+1),\\
&& \frac{d\rho_{de}}{dX}=\rho^{0}_{cr}\left[\nu_{m}-\nu_{G} (1-\nu_{m}) \ln{X}\right],
\eea whose solutions are given by
\bea
&& \rho_{m}(X)=\rho^{0}_{m}+\rho^{0}_{cr}\left[-1+\nu_{m}+\nu_{G}\left(1-\nu_{m}\right)\right]+\nonumber\\
&& X \rho^{0}_{cr} \left[1-\nu_{m}-\nu_{G}\left(1-\nu_{m}\right)\left(1-\ln X\right)\right],
\eea and 
\bea
&& \rho_{de}(X)=\rho^{0}_{de}-\rho^{0}_{cr}\left[\nu_{m}+\nu_{G}\left(1-\nu_{m}\right)\right]+\nonumber\\
&& X \rho^{0}_{cr} \left[\nu_{m}+\nu_{G} \left(1-\nu_{m}\right) \left(1-\ln{X}\right)\right],
\eea respectively.

Upon replacement of the above solutions into Eq. \eqref{Eff_FR_Eq00}, one obtains $\rho^{0}_{cr}=\rho^{0}_{m}+\rho^{0}_{de}$. On the other hand, doing the same but now in Eq. \eqref{Eff_FR_Eqii} we find 
\be
\frac{dX}{dN}= \frac{3\left[\rho^{0}_{de}/\rho^{0}_{cr}-\nu_{m}- (1-\nu_{m}) \left(\nu_{G}+X \left(1-\nu_{G}\left(1-\ln X\right)\right)\right)\right]}{1+\nu_{G} \ln X}.
\label{dEdN}
\ee After solving this equation, we obtain the dimensionless Hubble rate squared $E^2(N)$, and then $H(N)$, or equivalently, $H(z)$, by introducing $N=-\ln(1+z)$. 

Also, from \eqref{Omega_de}, one has that 
\bea
&& \Omega_{de}(X)=\nu_{m}+\left[\nu_{G} (\nu_{m}-1)-\nu_{m}+\Omega^{0}_{de}\right]\frac{1}{X}+\nonumber\\
&&\nu_{G} (1-\nu_{m})\left(1-\ln{X}\right),
\label{Omega_de_Varying_G}
\eea and $\Omega_{m}(X)=1-\Omega_{de}(X)$.
From Eqs. \eqref{wde_G_Vary},\eqref{wT_G_Vary},  we find
\bea
&& w_{de}(X)=-\frac{1}{1+\nu_{G}\ln{X}}\Big[1+\nonumber\\
&& \frac{\nu_{G} X \ln{X}}{\nu_{G} ((1-\nu_{m})X+\nu_{m}-1)+X \nu_{G} (\nu_{m}-1) \ln{X}+(X-1) \nu_{m}+\Omega^{0}_{de}}\Big],
\label{w_de_Varying_G}
\eea and 
\be
w_{T}(X)=\frac{1}{(1+\nu_{G} \ln X)}\Bigg[ \left(1+\frac{1}{X}\right)\left[\nu_{m}+\nu_{G}\left(1-\nu_{m}\right)\right]-\frac{\Omega^{0}_{de}}{X}-\nu_{G}\nu_{m} \ln{X}\Bigg].
\label{w_T_de_Varying_G}
\ee 

The equation \eqref{dEdN} is an autonomous equation which can be treated following the same framework of dynamical systems. This equation has a single critical point $X_{c}$ which can be determined from 
\be
\ln X_{c}= \frac{X_{c} (\nu_{G}-1) (\nu_{m}-1)-\nu_{G}\nu_{m}+\nu_{G}+\nu_{m}-\Omega^{0}_{de}}{X_{c}\nu_{G} (\nu_{m}-1)}.
\ee  For example, numerically we have found that for $\Omega^{0}_{de}\simeq 0.73$, $\nu_{dm}\simeq 0.01$ and $\nu_{G}\simeq 1\times 10^{-3}$, the value $X_{c}$ is approximately $X_{c}\simeq 0.73$. This result gives us the asymptotic (future) value of the Hubble rate by using the relation $H_{c}=H_{0} \sqrt{X_{c}}$. So, from the observational data of Planck $2018$ Ref. \cite{planck2}, one has that $H_{0}=67.4$ Km/(Mpc sec) and therefore $H_{c}\simeq 57.48$ Km/(Mpc sec) at $z\rightarrow -1$.

By substituting in Eqs. \eqref{Omega_de_Varying_G},\eqref{Omega_de_Varying_G}, and \eqref{Omega_de_Varying_G}, one can see that it is a  Sitter solution representing dark energy 
dominance with $\Omega_{de}=1$, $w_{de}=-1$, and $w_{T}=-1$. 

The stability of this fixed point can be studied by considering the solution
\be
X(N)=X_{c}+\delta{X(N)},
\ee where the perturbation $\delta{X(N)}$ satisfies $\delta{X(N)}\ll 1$. Thus, replacing it in Eq. \eqref{dEdN} we obtain
\be
\frac{d\delta{X}}{dN}=-3 (1 - \nu_{m}) \delta{X},
\ee whose solution is 
\be
\delta{X}= C e^{\mu N},
\ee with $\mu=-3 (1-\nu_{m}) $. So, since $\nu_{m}\ll 1$ then $\mu<0$ and accordingly, the fixed point is always an attractor. We note that the stability does not depend on $\nu_{G}$ at leading order in perturbation. 

Similarly, by using Eqs. \eqref{rz} and \eqref{rsz}, the statefinder parameters $r$ and $s$ are computed to be 
\bea
&& r=1-\frac{9 \nu_{m} (1-\Omega^{0}_{de}) (z+1)^{3\left(1- \nu_{m}\right)}}{2 X\left(1+\nu_{G}\ln X\right)}-\frac{9 \nu_{G} (1-\Omega^{0}_{de})^2 (z+1)^{6\left(1-\nu_{m}\right)}}{2 X^2 (1+\nu_{G}\ln X)^3},\\
&& s=\frac{(1-\Omega^{0}_{de})\left[\nu_{m} X (1+\nu_{G} \ln X)^2+\nu_{G} (1-\Omega^{0}_{de}) (z+1)^{3\left(1-\nu_{m}\right)}\right]}{X (1+\nu_{G} \ln X)^2 \left[-1+\Omega^{0}_{de}+X (z+1)^{3\left(\nu_{m}-1\right)} (1+\nu_{G}\ln X)\right]},
\eea
which reduce to those already obtained for Model II in the limit $\nu_{G}\rightarrow 0$.

\begin{figure*}[h!]
\centering
\includegraphics[width=0.48\textwidth]{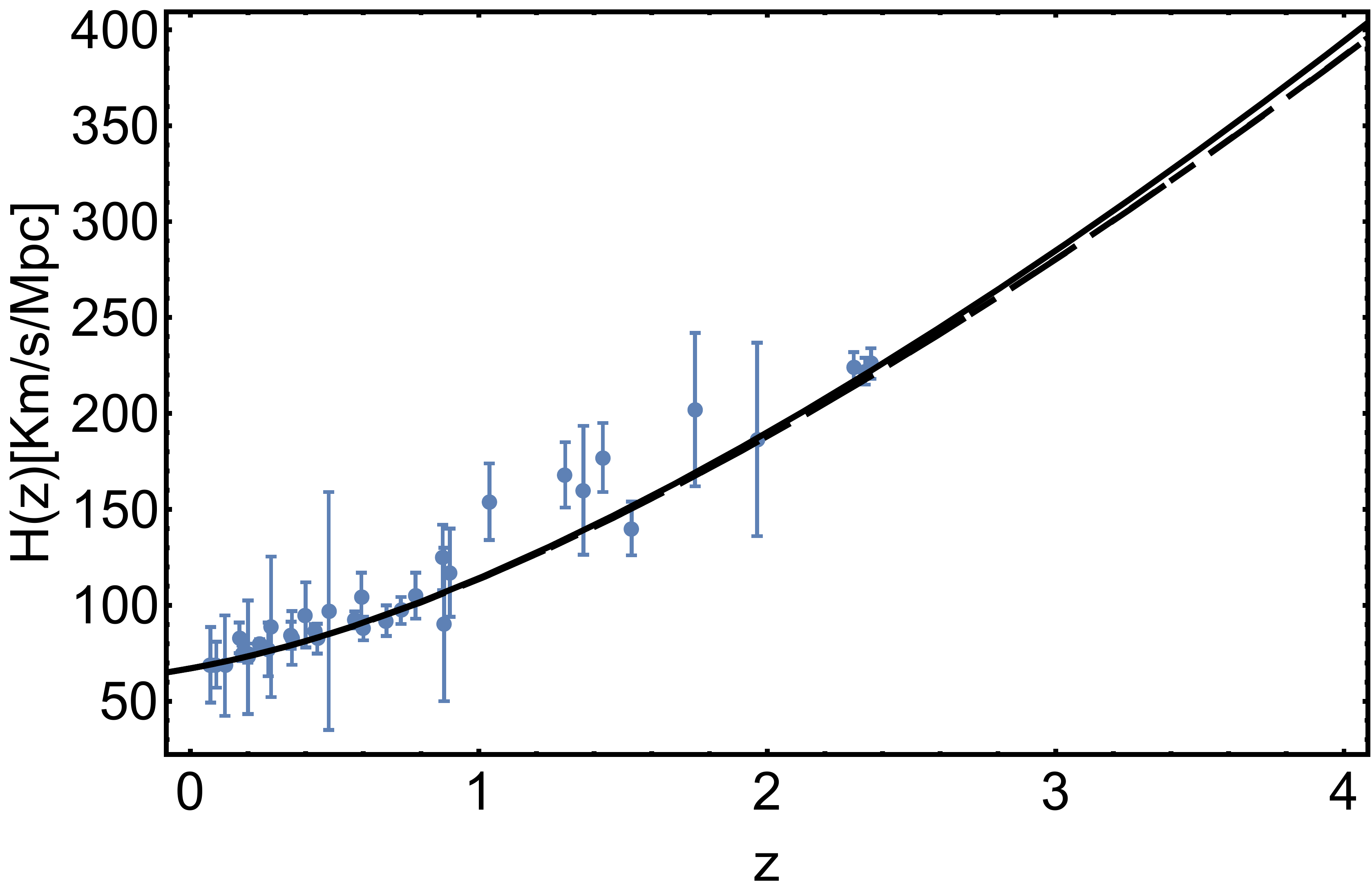} 
\includegraphics[width=0.48\textwidth]{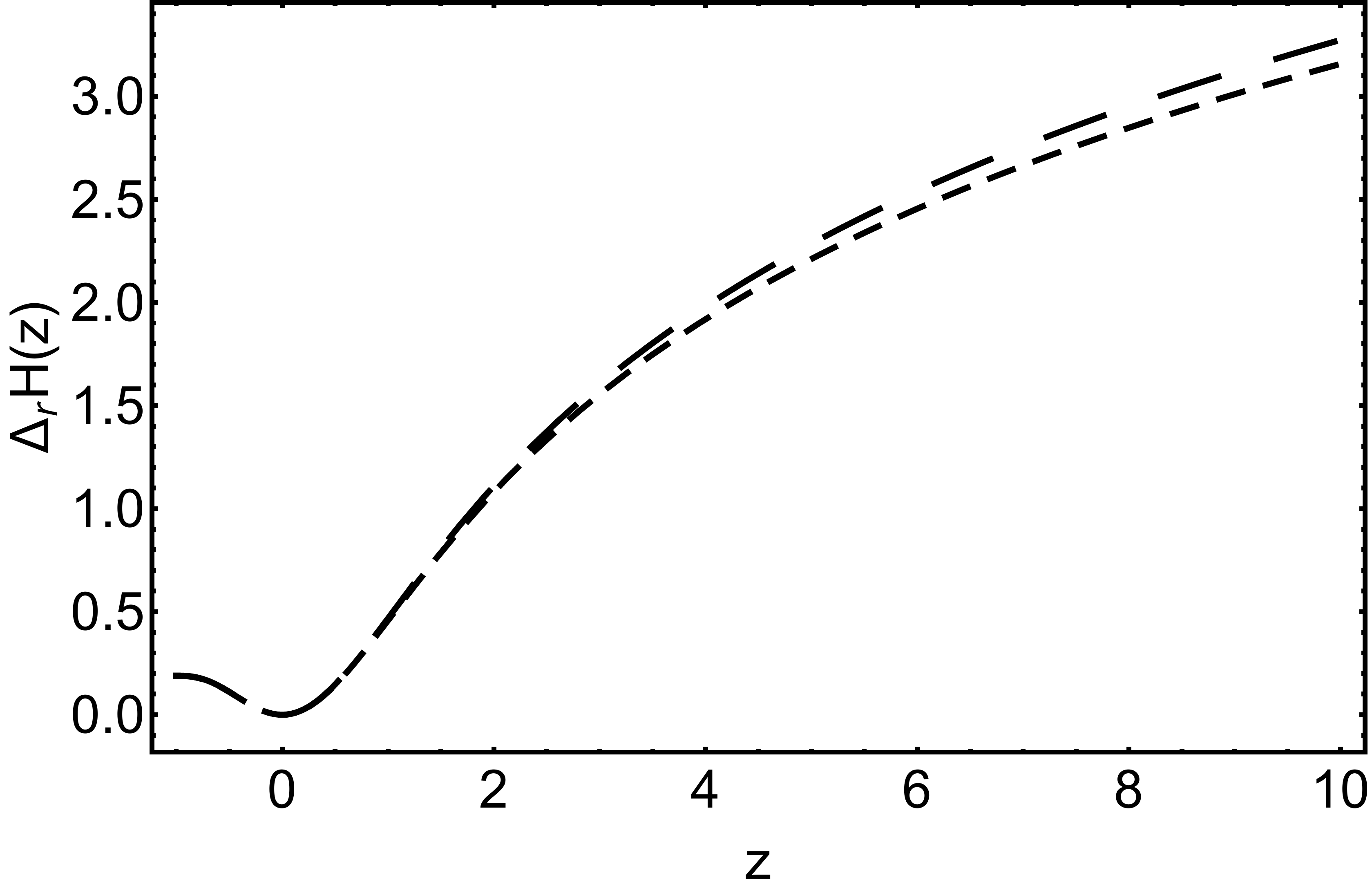}
\caption{
It is shown the evolution of the Hubble rate $H(z)$ (left panel) as a function of red-shift $z$, for the $G$-varying model defined in Eq. \eqref{G_H}, with the coupling function  $Q=3 \nu_{dm}H\rho_{dm}$, representing the interaction between dark energy and dark matter, for $\nu_{m}=0.01$ and $\nu_{G}=1\times 10^{-3}$ (dashed line), as also, the Hubble rate of $\Lambda$CDM model, along with the Hubble data from Refs. \cite{Meng:2015loa}, and \cite{Farooq:2013hq}. In the right panel it is depicted the behaviour of the exact relative difference (percentage) defined as $\Delta_{r}{H}(z)\equiv 100\times |H-H_{\Lambda CDM}|/H_{\Lambda CDM}$ with respect to the concordance model, for $\nu_{dm}=0.01$ and two different values of $\nu_{G}$, the first value $\nu_{G}=5\times 10^{-4}$ (short-dashed line) and the second one $\nu_{G}=1\times 10^{-3}$ (large-dashed line). We have used  $H_{0}=67.4$ Km/(Mpc sec) from Planck $2018$ \cite{planck2}.
}
\label{fig:12}
\end{figure*}

In Fig.~\ref{fig:12} (left panel), it is shown the evolution of the Hubble rate $H(z)$ for the present model by solving the differential equation \eqref{dEdN} for some values of $\nu_{dm}$ and $\nu_{G}$. It is also added the Hubble rate $H_{\Lambda CDM}$ of the $\Lambda$CDM model, along with the current available data for $H(z)$ from \cite{Meng:2015loa} and \cite{Farooq:2013hq}. Also, we depict the behaviour of the exact relative difference with respect to the concordance model, for fixed $\nu_{dm}$ and a pair of different values of $\nu_{G}$. We take $\nu_{dm}=0.01$ and two different values of $\nu_{G}$, the first value $\nu_{G}=5\times 10^{-4}$ and the second one $\nu_{G}=1\times 10^{-3}$, which are included into the physical range $\nu_{G} \in \left[5\times 10^{-4},1\times 10^{-3}\right]$, obtained from observations, see for example Refs. \cite{sola1,sola3,Sola:2016ecz}. It is observed an increasing of $\Delta_{r}{H}$ for higher red-shifts, $z\gtrsim 2$, and after the present time $z=0$, in the future. Particularly, for $z\simeq 10$ we obtain $\Delta_{r}{H}\simeq 3 \%$, whereas that for $z=-1$ we have $\Delta_{r}{H}\simeq 0.18\%$.

\begin{figure*}[h!]
\centering
\includegraphics[width=0.48\textwidth]{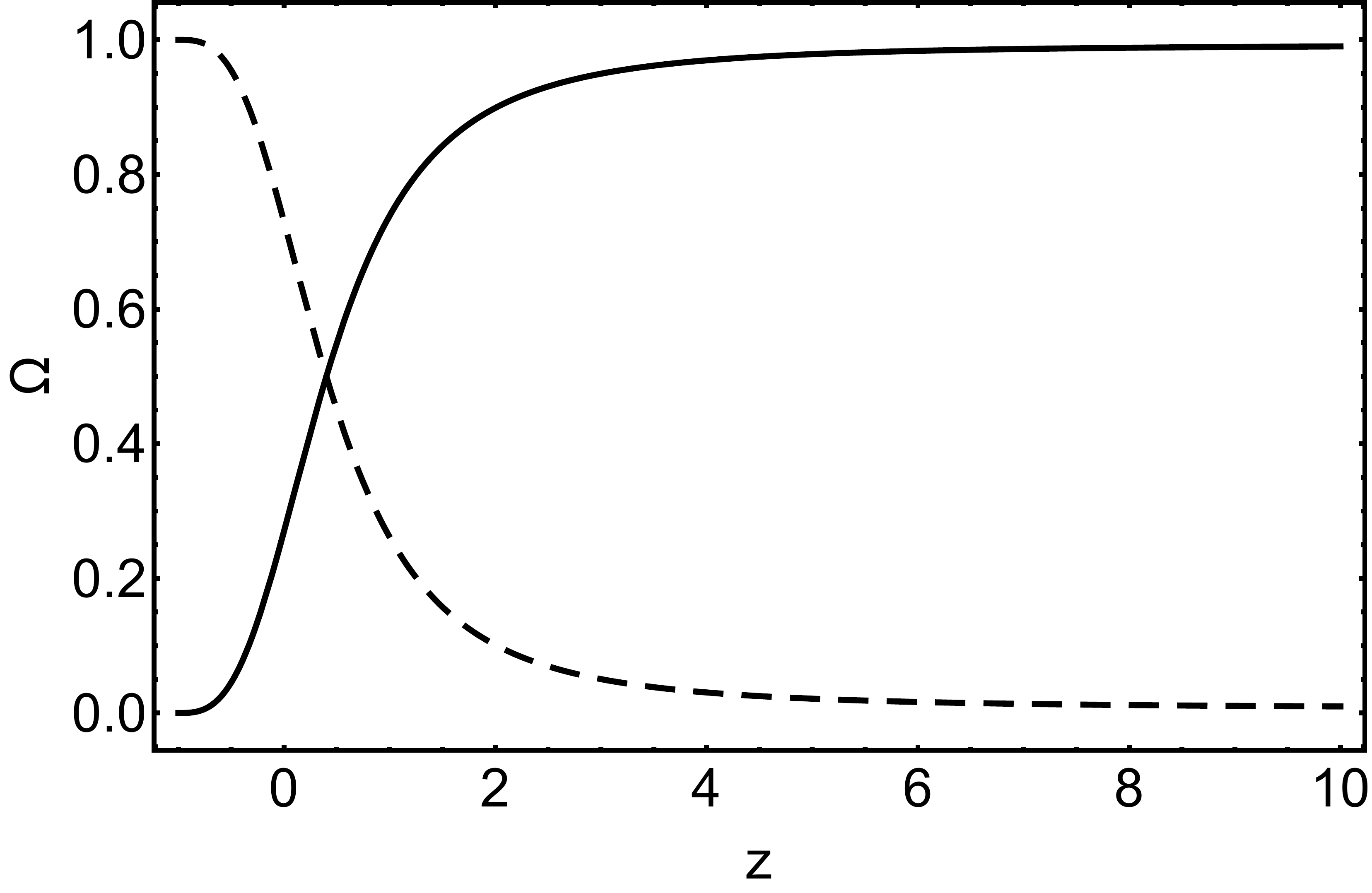} 
\includegraphics[width=0.48\textwidth]{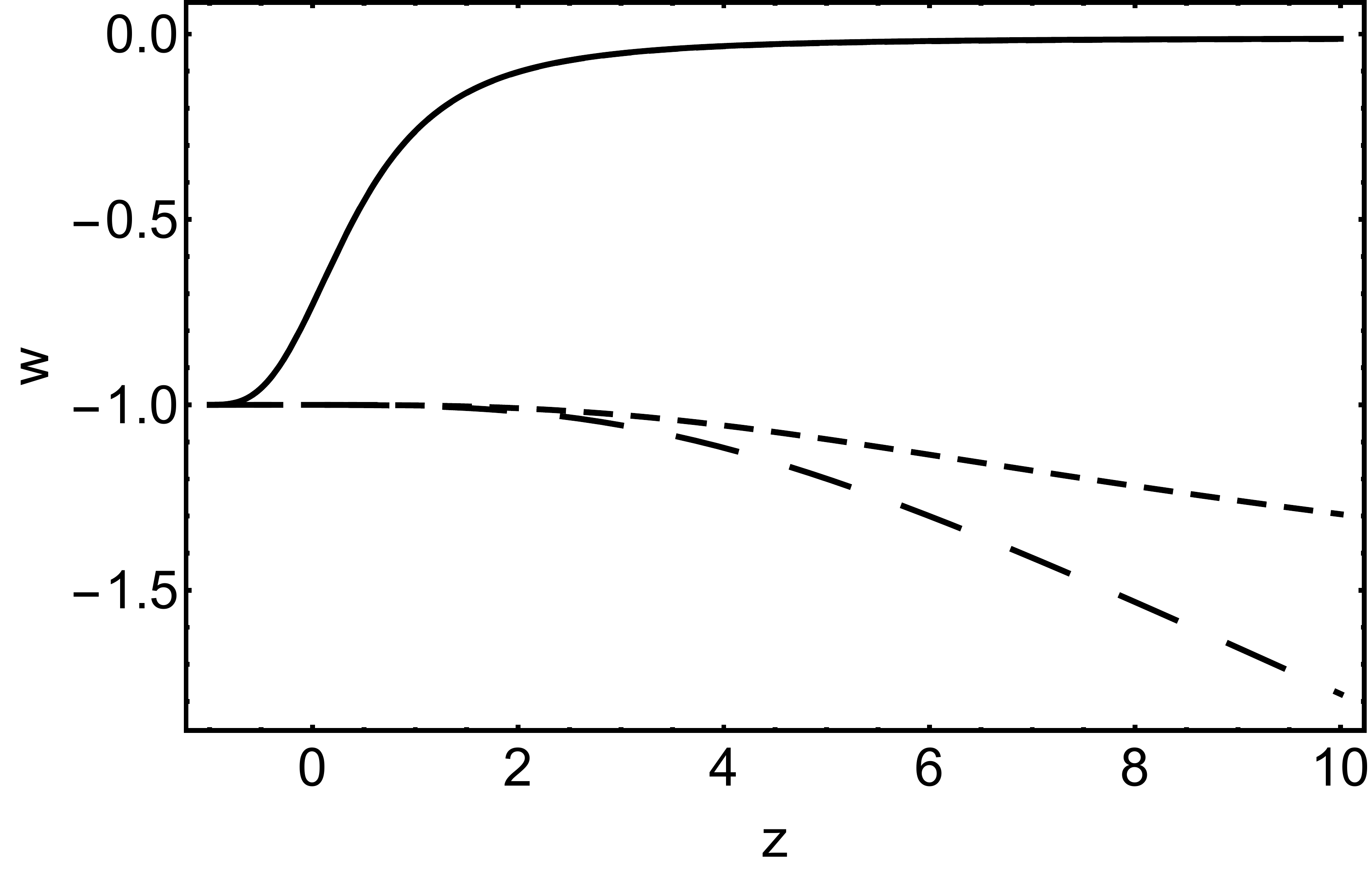}
\caption{
In the left panel we depict the behaviour of the fractional energy densities of dark energy $\Omega_{de}$ (solid line) and cold dark matter $\Omega_{dm}$ (dashed line), as functions of redshift $z$, for $\nu_{m}=0.01$ and $\nu_{G}=5\times 10^{-4}$. In the right panel we show the behaviour of the EOS parameter of dark energy $w_{de}$ (short-dashed line) and the total EOS parameter $w_{T}$ (solid line) as functions of $z$, for $\nu_{m}=0.01$ and $\nu_{G}=5\times 10^{-4}$. Also, the large-dashed line corresponds to values of $w_{de}$, but for a larger value of $\nu_{G}$, in such a way that we now have $\nu_{G}=1\times 10^{-3}$. At the present time, at $z=0$, they take the values $\Omega_{de}\simeq 0.73$, $\Omega_{dm}=0.27$, $w_{de}\simeq -1$, and $w_{T}\simeq -1$. Asymptotically, the model approaches to an attractor fixed point which is a de Sitter dark energy dominated solution with $\Omega_{de}=1$, $\Omega_{dm}=0$, $w_{de}= -1$, and $w_{T}= -1$. 
}
\label{fig:9}
\end{figure*}

In Fig.~\ref{fig:9} we plot the variation with respect to $z$ of some cosmological parameters such the fractional energy densities of dark energy $\Omega_{de}(z)$, dark matter $\Omega_{dm}(z)$, the EOS parameter of dark energy $w_{de}(z)$, and the total EOS parameter $w_{T}(z)$. It is seen that the model can explain the current accelerated expansion of the universe and at $z=0$ it gives $\Omega^{0}_{de}\simeq 0.73$, $\Omega_{dm}^{0}\simeq 0.27$, $w^{0}_{de}\simeq -1$, and $w^{0}_{T}\simeq -0.73$. Also, when $z\rightarrow -1$, the model tends asymptotically to an attractor which is a de Sitter solution with $\Omega_{de}=1$, $\Omega_{dm}=0$, $w_{de}=-1$ and $w_{T}=-1$. For higher red-shifts we observe that $w_{de}$ becomes more sensitive to the values of $\nu_{G}$, taking smaller values than $-1$ and going deeper in the phantom regime for larger values of $\nu_{G}$. Nevertheless, let us note that the energy density of dark energy decays very quickly and the effective cosmic fluid behaves as nonrelativistic matter with $w_{T}\simeq 0$, and therefore allowing the existence of the standard matter-dominated era \cite{copeland, Gonzalez-Espinoza:2018gyl}.

\begin{figure*}[h!]
\centering
\includegraphics[width=0.48\textwidth]{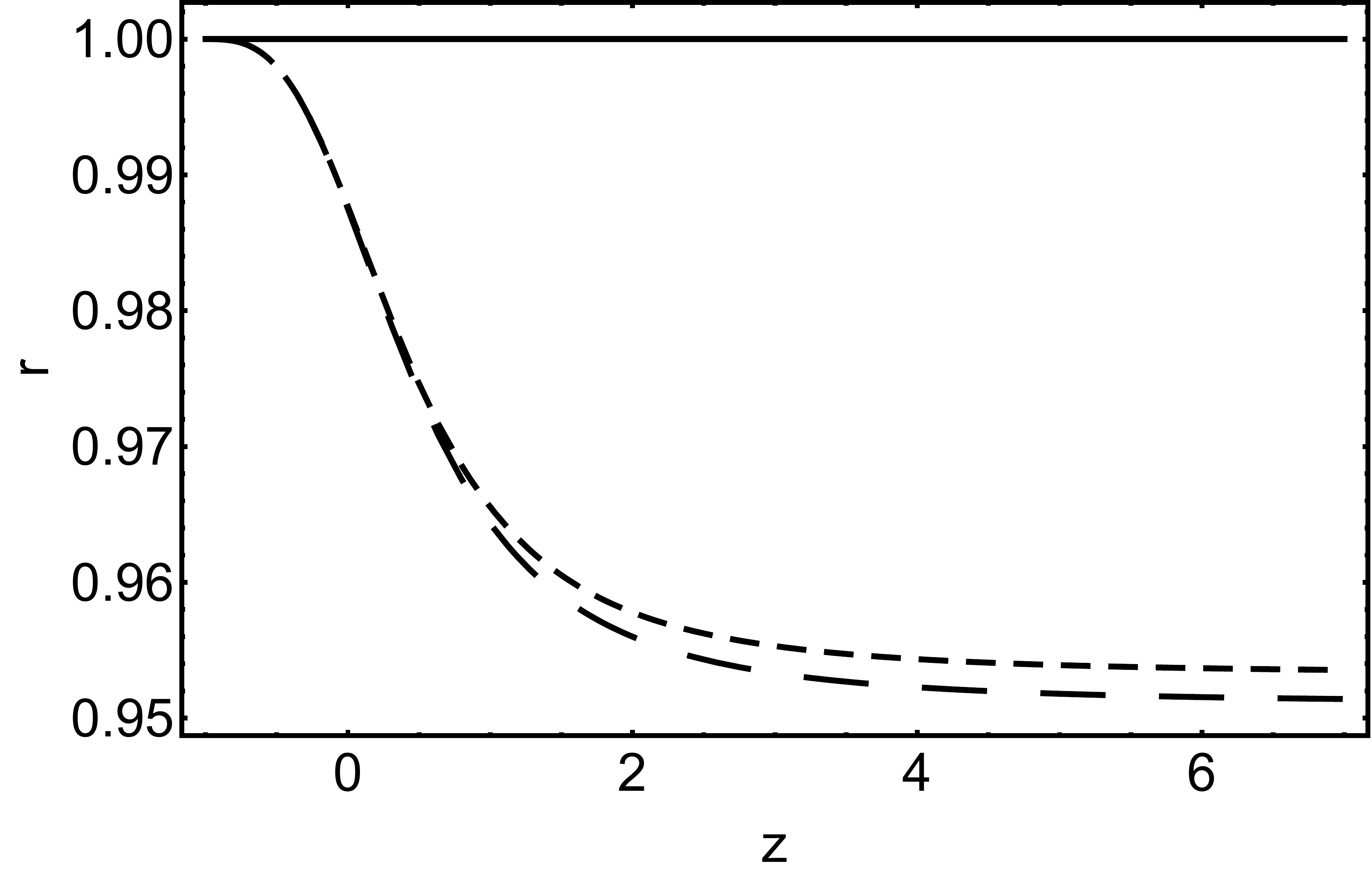} 
\includegraphics[width=0.48\textwidth]{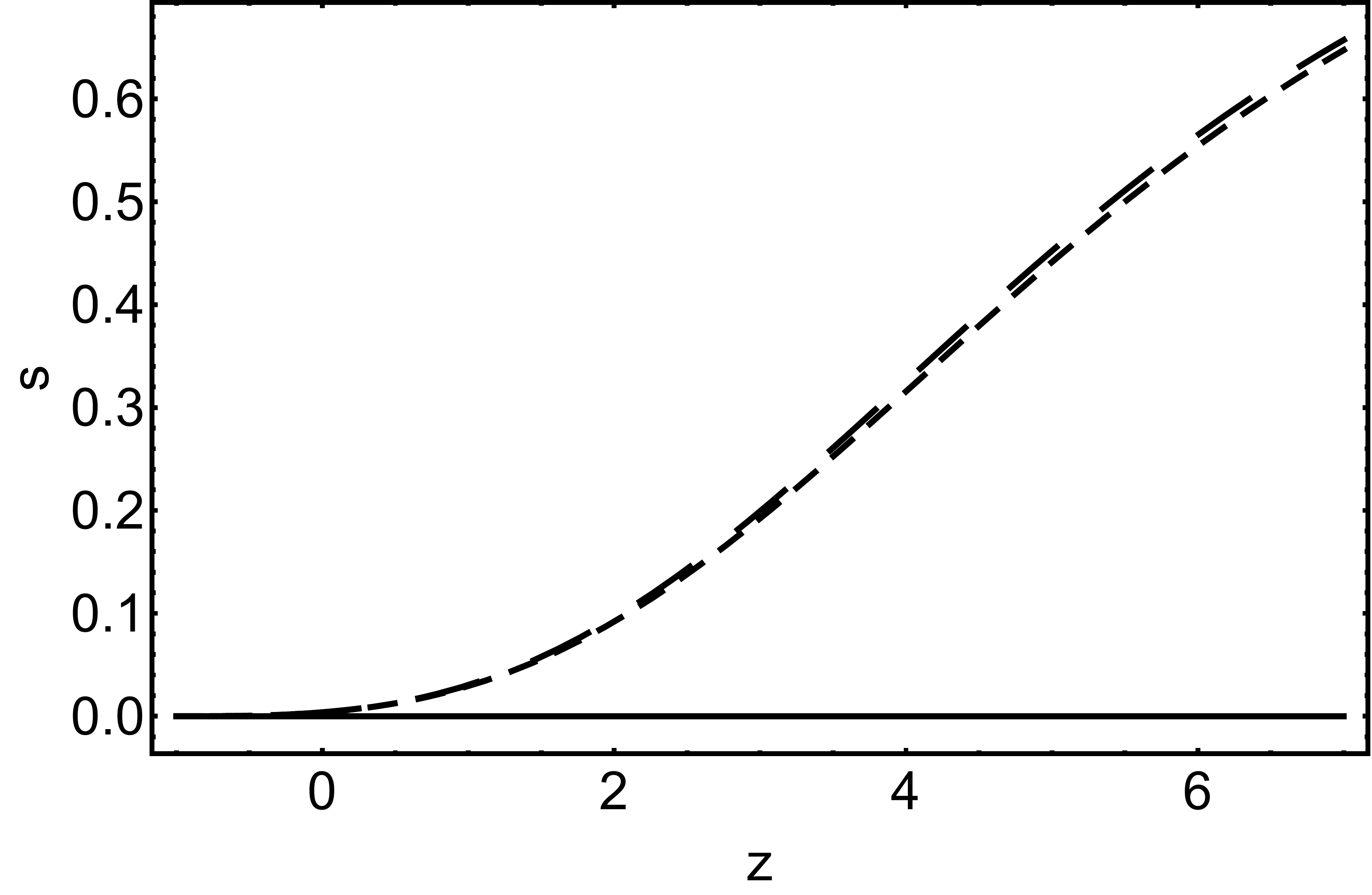}
\caption{
It is depicted the behaviour of the statefinder parameters $r$ (left panel) and $s$ (right panel) as functions of redshift $z$, for $\nu_{m}=0.01$, and $\nu_{G}=5\times 10^{-4}$ (short-dashed line) and the second $\nu_{G}=1 \times 10^{-3}$ (large-dashed line). Solid line represents the values of $r$ and $s$ for $\Lambda$CDM model. It is seen that asymptotically, when $z\rightarrow -1$, the system tends to a de Sitter solution, $r=1$ and $s=0$, consistently with previous results of dynamical systems.
}
\label{fig:11}
\end{figure*}

Finally, in Fig.~\ref{fig:11} we show the evolution of the statefinder parameters $r(z)$ (left panel) and $s(z)$ (right panel) as functions of red-shift, for the same set of values of parameters used in above plots. It can be seen that at $z=0$, and for $\nu_{dm}=0.01$, $\nu_{G}=5\times 10^{-4}$ (short dashed line), these parameters take the values $r\simeq 0.988 $ and $s\simeq 3.75\times 10^{-3}$. For the larger value $\nu_{G}=1\times 10^{-3}$ (large dashed line), at $z=0$, we get $r\simeq 0.988 $ and $s\simeq 3.80 \times 10^{-3}$. So, for lower red-shift, there is a small difference (of the order of $4\%$ for $z\lesssim 2$) between the results when varying $\nu_{G}$, and this difference is much smaller for the values of $r$ than for $s$. When $z\rightarrow -1$, the trajectory of the system in the plane of statefinder parameters tends toward the de Sitter expansion at the future, with $r=1$ and $s=0$. 

%\begin{figure*}[h!]
%\centering
%\includegraphics[width=0.48\textwidth]{G_Varying_model_G.pdf} 
%\includegraphics[width=0.48\textwidth]{G_Varying_model_dGdz.pdf}
%\caption{In the left panel we show the evolution of the gravitational coupling $G(z)/G_{0}$ as function of the redshift for same initial conditions of the above plot. In particular, we have $\nu_{m}=0.01$, and take two different values of $\nu_{G}$, the first one is $\nu_{G}=5\times 10^{-4}$ (short-dashed line) and the second $\nu_{G}=1 \times 10^{-3}$ (large-dashed line). On the other hand, in the right panel it is shown the evolution of the cosmic drift rate $\dot{G}/G$ as a function of $z$, also for the same values of parameters and initial conditions.}
%\label{fig:10}
%\end{figure*}

\newpage

%%%%%%%%%%%%%%%%%%%%%%%%
\section{Concluding Remarks}
%%%%%%%%%%%%%%%%%%%%%%%

In summary, in the present work we have applied phase-space dynamical techniques to three running vacuum dark energy models, and we have computed the statefinder parameters as functions of the red-shift.

From our dynamical analysis we have shown that the two models, I, and II can explain the current accelerated expansion phase of the Universe, being that the corresponding fixed point, either a de Sitter solution (Model I) or scaling solution (Model II), is also an attractor in all the cases, provided that $\nu_{i}=\{\nu_{\text{dm}},\nu_{\Lambda}\}<1$. Also, for Model I and II the thermal history of the universe can be successfully reproduced, from the radiation-dominated era, passing through the matter-dominated era, and finally reaching the dark energy-dominated phase. 

More interestingly, in the case of Model I, the fixed point representing the matter era is a scaling solution in which the dark energy density has a contribution to the total energy density during the dark matter era, with $\Omega_{\Lambda}=\nu_{\text{dm}}$. So, when the parameter $\nu_{\text{dm}}$ is small this contribution is also small. Thus, the parameter $\nu_{\text{dm}}$ indicates a slight deviation with respect to the standard mater era whose value may be more constrained from large-scale structure (LSS) data. It is due to the fact that when the universe enters in this fixed point the growing of matter density perturbations can be suppressed by the presence of dark energy, such that the dark matter density contrast grows less rapidly that the first power of the scale factor, depending on the amount of dark energy \cite{Amendola:1999er}. On the other hand, in the case of model II, the final attractor is a scaling solution with accelerated expansion which satisfies $\Omega_{\Lambda}=1-\nu_{\Lambda}$, $w_{DE}=-1$ and $w_{T}=-1+\nu_{\Lambda}$. This class of solution is very interesting because it provides a natural mechanism in alleviating the fine-tuning problem, or cosmological coincidence problem of dark energy \cite{copeland}. It can adjust the current values of the cosmological parameters such as $\Omega_{\Lambda}^{0}=0.7$ and $\Omega_{m}^{0}=0.3$, and at the same time explain the accelerated expansion. 

Regarding the statefinder analysis, first let us recall that the pair $\{r,s\}$ is defined using third order time derivatives of the scale factor, and that the statefinder parameters have the potential to discriminate between different dark energy models.
%we should point out that the parameter $r$ is considered as the next step in the hierarchy of geometrical cosmological parameters after the Hubble parameter $H$ and the deceleration parameter $q$. Similarly, the second statefinder parameter $s$ is a linear combination of $q$ and $r$ written conveniently, such as this quantity does not depend on dark energy density. Thus, this pair $\{r,s\}$ effectively allow us to differentiate between several DE models. 
Now, we should discriminate between the two cases shown in Fig.~\ref{fig:3}. The first column was plotted for $\nu_{i}= 0.01$, the second column was depicted for $\nu_{i}=0.001$. Starting from the left panel, we observe evident discrepancies between the models. This difference is also natural because running vacuum models are quantum-inspired, which means that the deviations with respect to the classical counterpart should be, in general, small. Taking the above idea seriously, the parameter $\nu_{i}$, which encodes the quantum features, should be taken in such a way that the effect on the classical solution will be soft. We then claim that the last column in Fig.~\ref{fig:3} should be taken as a more suitable situation.
Furthermore, notice that when $\nu_{i}$ is taken to be close to zero (second column), the deceleration parameter $q(z)$ looks qualitative identical to those for $\Lambda$CDM. 
If we now analyse the parameter $r$, we observe once more a notorious similarity to the standard scenario, although the second parameter (i.e., $s$) exhibits a remarkable difference. Thus, although classically these models should be equivalent, at the level of the statefinder diagnostic, this is not the case.

Finally, it has been performed a further analysis regarding to a class of models with variable gravitational coupling within the interacting vacuum scenario, by using the tools dynamical system and statefinder analysis. In doing so, we have studied a particular model for $G(H)$ from Ref.  \cite{Fritzsch:2016ewd}, and $Q=3 \nu_{dm} H \rho_{dm}$. It has been introduced an effective cosmic fluid for describing DE by defining, both effective energy density and pressure, which contain the contribution of the running of gravitational coupling. We have shown that the model has only fixed point which is an attractor and de Sitter solution, allowing to adjust the current data of $H(z)$, along with the other cosmological parameters such that the fractional energy density of DE and the equation of state of dark energy at the present time. In particular,  we found  that the effect of $\nu_{dm}$ becomes more significant for higher redshift, and at the future when $z\rightarrow -1$ in comparison with the $\Lambda$CDM model. Regarding the statefinder analysis we observe that the $s$ parameter becomes more sensitive to higher redshifts than the $r$ parameter, when we vary the  coupling$\nu_{G}$.
As a final remark, an exhaustive study including several anszats 
for both the gravitational coupling and the interaction between DE and DM deserves a separated project, reason why these ideas will de addressed in a future work.

\bigskip

{\bf Note added:} One day before we received the referee report, a new work related to ours appeared \cite{Extra_Ref}, which indicates that the topic is interesting. Our work was carried out in complete independence from them, and vice versa. In that work, the authors have analysed the dynamics and evolution of several $\Lambda$-varying cosmological models. The critical points and their nature have been determined, and the corresponding phase space is shown, although they have not discussed at all the statefinder parameters. We have checked that where there is overlap their results are in agreement with ours, which is a confirmation that both calculations are error-free.
%%%%%%%%%%%%%%%%%%%%%%%%%%%%%%%%%%%%%%%%%%%%%%%%%%%%%%%%%%%%%%%%%%%%%%%%%%%%%%%%%%%%%

\begin{acknowledgments}
We are grateful to the anonymous reviewer for a careful reading of the manuscript, 
for her/his constructive criticism and for valuable comments and suggestions.  
The author G.~P. thanks the Funda\c c\~ao para a Ci\^encia e Tecnologia (FCT), 
Portugal, for the financial support to the Center for Astrophysics and Gravitation-CENTRA,  
Instituto Superior T\'ecnico, Universidade de Lisboa, through the Grant No. UID/FIS/00099/2013. 
The author \'A.~R. acknowledges DI-VRIEA for financial support through Proyecto Postdoctorado 2019 VRIEA-PUCV. The author N.~V. was supported by Comisi\'on Nacional de Ciencias y Tecnolog\'ia of Chile through FONDECYT Grant N$^{\textup{o}}$ 11170162. Additionally, N.~V. would like to express his gratitude to the Instituto Superior T \'ecnico of Universidade de Lisboa for its kind hospitality during the final stages of this work. The author G.~O acknowldeges DI-VRIEA for financial support through Proyecto Postdoctorado $2019$ VRIEA-PUCV.
\end{acknowledgments}

%%%%%%%%%%%%%%%%%%%%%%%%%%%%%%%%%%%%%%%%%%%%%%%%%%%%%%%%%%%%%%%%%%%%%%%%%%%%%%%%%%%%%

\end{document}